\documentstyle[multicol,aps,amssymb,epsfig]{revtex}

\begin{document}

\title{On the Extension Behavior of Helicogenic Polypeptides}
\author{Arnaud Buhot}
\address{Theoretical Physics, University of Oxford, 1 Keble Road,\\
Oxford, OX1 3NP, UK}
\author{Avraham Halperin}
\address{UMR 5819 (CEA, CNRS, UJF), DRFMC/SI3M, \\
CEA-Grenoble, \\
17 rue des Martyrs, 38054 Grenoble Cedex 9, France}
\date{\today }
\maketitle

\begin{abstract}
The force laws governing the extension behavior of homopolypeptides are
obtained from a phenomenological free energy capable of describing the
helix-coil transition. Just above the melting temperature of the free
chains, $T^{\ast }$, the plot of force, $f$, {\it vs.} end-to-end 
distance, $R$,
exhibits two plateaus associated with coexistence of helical and coil
domains. The lower plateau is due to tension induced onset of helix-coil
transition. The higher plateau corresponds to the melting of the helices by
overextension. Just below $T^{\ast }$ the $fR$ plot exhibits only the upper
plateau. The $fR$ plots, the helical fraction, the number of domains and
their polydispersity are calculated for two models: In one the helical
domains are viewed as rigid rods while in the second they are treated as
worm like chains. 
\end{abstract}

\pacs{PACS numbers: 61.25.Hq, 61.41.+e, 87.15.He}

\begin{multicols}{2}
\narrowtext

\section{Introduction}

Single molecule biomechanical experiments allow to measure the forces
generated by biomolecules and their response to applied forces~\cite{Mehta}.
For ``passive'' biopolymers, that do not produce active movement against
load, the resulting force-extension curves provide a probe of the nonlinear
elasticity of the chains and its relationship to configurational changes.
Such studies were carried out for single stranded and double stranded 
DNA~\cite{Bustamente,Bensimon,Austin,Bensimon2}, for the muscle protein 
Titin~\cite{Rief,Kellermayer,Tskhovrebova}, the extracellular matrix 
protein Tenascin~\cite{Oberhausser}, the polysaccharides 
Dextran~\cite{Rief2} and Xanthan~\cite{Li}, as well as the synthetic 
polymer poly(ethylene-glycol)~\cite{PEG}. Theoretical modeling of the 
elastic behavior of these polymers
is often hampered by lack of detailed knowledge concerning the
configurations involved and the relevant interactions. A direct and detailed
confrontation between experiment and theory is however possible for the case
of homopolypeptides capable of forming an $\alpha $-helix. In this case the
molecular configurations are well understood. Furthermore, one can relate
the elastic force law to the known Zimm-Bragg parameters that characterize
the helix-coil transition of the free polypeptides. As we shall discuss, the
extension force law of a long polypeptide just above the melting temperature
of the helix, $T^{\ast }$, exhibits two plateaus. Both are associated with a
coexistence of helical and coil domains. The low-tension plateau is due to
helix formation induced by the chain extension and the accompanying loss of
configurational entropy. The second, high-tension plateau is due to the
extension induced melting of the helices when the end-to-end length exceeds
the length of the fully helical chain. For $T<T^{\ast }$ only the second
plateau survives but the force law for weak extensions exhibits a ``quasi
plateau'' {\em i.e.,} a regime with a weak slope that is not associated with
a coexistence. This last regime is due to the facile alignment of the long
persistent regions in the helix. In the following we analyze the force laws,
their dependence on temperature and the corresponding population
distribution of helical and coil segments. While we are unaware of single
molecule experiments on this system, similar behavior was observed
experimentally on fibers of Collagen during the 1950s~\cite{Flory}.

\begin{figure}[tbp]
\begin{center}
\epsfig{file=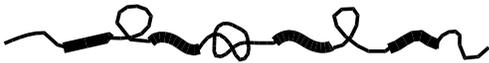,width=3in}
\caption{\label{multi} Schematic picture of the 
stretching of a polypeptide comprizing different 
helical domains ({\it thick lines})
and coil domains ({\it thin lines}).}
\end{center}
\end{figure}

Two different treatments of this problem were recently 
advanced~\cite{Buhot,Mario}. In the present article we present a unified 
analysis of the
problem, tracing the physical origins of the disagreements between the force
laws obtained in the two earlier publications and analyzing their range of
validity. Our treatment is distinctive in that it is based on free energy
argument instead of the customary transfer matrix 
approach~\cite{Birshtein,Poland,Grosberg}. 
This free energy argument allows to recover the
results of the transfer matrix method but is physically transparent and
relatively simple mathematically. Within this approach we consider the
extension of a very long homopolypeptide capable of forming 
$\alpha $-helices. The discussion is limited to quasi-static extension 
assuming that
the rate of equilibration of the configurations of the chain is much faster
than the rate of extension. Two models are explored. In one, the helical
domains are viewed as rigid rods while in the second, they are modeled as
semiflexible chains. For each model we present a rigorous derivation of the
corresponding force law allowing for the polydispersity of the helical and
coil domains and the associated mixing entropy (see Fig.~\ref{multi}).
In addition, we explore two useful approximations allowing for simplified 
calculations of the force law. In one approximation, the mixing entropy 
of the helices and domains is neglected and the plateau is associated 
with coexistence of one helical domain with one coil domain (see 
Fig.~\ref{phases}). Within this ``diblock'' or ``$S_{mix}=0$''
approximation the coexistence regimes involve a first order phase
transition. In spite of this erroneous conclusion, this approximation
correctly identifies the important length and force scales in the problem.
The second, mean field approximation, neglects the direct coupling of the
distribution of domains sizes and the applied tension. As we shall see, this
approximation is actually exact when the helical domains are modeled as long
semiflexible chains.

The problem and the analysis are of interest from a number of perspectives.
Homopolypeptides are the simplest representatives of helicogenic molecules.
In this case it is not necessary to allow for the sequence heterogeneity
which affects the elasticity of heteropolypeptides. Nor is it necessary to
consider the long range interaction that occur in multiple helices. As a
result, the number of parameters invoked in the theory is smaller and the
mathematical analysis is simpler. The force laws obtained are determined by
known Zimm-Bragg parameters with no adjustable parameters. Accordingly this
system enables a direct comparison between experiment and theory. In this
juncture it is important to note the availability of synthetic methods for
producing long polypeptides with a well defined architecture~\cite{Deming}.
Furthermore, the formalism described provides a convenient basis for the
analysis of more complex helicogenic polymers such as DNA and Collagen. From
a polymer science point of view the elasticity of homopolypeptides is of
interest because it reflects the effects of internal degrees of freedom
arising from intrachain self-assembly. The study of these systems enables
the exploration of the distinctive non-linear force laws that are the
signatures of such intrachain self-assembly. Finally, this elasticity plays
a role in the stabilization of the helical state of grafted 
polypeptides~\cite{Buhot2}. In turn this is of importance for the function 
of fusogenic
polypeptides and the design of surfaces that favor spreading and growth of
cells~\cite{Yu,Fields,James,Tjia}.

The paper is organized as follows. In section II we introduce the free
energy argument {\em via} an analysis of the helix-coil transition in free,
undeformed homopolypeptides. A semiquantitative analysis of the effect of
extension is presented in section III. Two approaches are used. One focuses
on the intersections of the free energy curves of a pure coil and a pure 
helix as a function of the end-to-end distance, $R$. As we shall see, each
intersection is a rough diagnostic for a transition between the two
configurations and for the occurrence of a plateau in the force law. The
second, more quantitative approach is the diblock or $S_{mix}=0$
approximation. Within this approximation the homopolypeptide is assumed to
consist of two domains, a helix and a coil, and the mixing entropy
associated with the polydispersity of the helical and coil domains is set to
zero, $S_{mix}=0$. Finally, in section IV we present a rigorous analysis of
the force laws allowing for the polydispersity of the coexisting helices and
coil domains. As stated earlier, we implement this analysis for two models.
In one the helical domains are viewed as rigid rods and in the second as
semiflexible chains described by the worm like chain (WLC) model. In this
section we also examine the validity of the mean field approximation
introduced in~\cite{Buhot}. Within this approximation, the explicit coupling
between the tension and the distribution of domain sizes is ignored. As we
shall see, it recovers the exact result when the helices are semiflexible 
{\em i.e.,} when the persistence length of the helices is finite and the
helical domains are large. The analysis of the rigid helices model recovers
the results obtained by Tamashiro and Pincus~\cite{Mario} using a transfer
matrix formalism. A comparison between the two models and an outline of some
open problems are presented in the Discussion.

\section{Helix-coil transition}

It is helpful to first summarize the main features of the helix-coil
transition of free polypeptides in the absence of imposed extension. We will
consider only homopolypeptides, consisting of a single type of amino acid
residues, in order to avoid complications introduced by sequence of
different residues found in heteropolypeptides. Our discussion focuses on
the case of long chains where the degree of polymerization of the polymer $N$
is large, $N\gg 1$. To this end we first discuss the Zimm-Brag parameters
and then introduce the free energy analysis of the transition. As we shall
see, this is equivalent to the transfer matrix formalism~\cite{footnote} 
that is traditionally used to analyze the 
transition~\cite{Birshtein,Poland,Grosberg}.

\subsection{Zimm-Bragg parameters}

Our discussion focuses on helicogenic polypeptides that are capable of
assuming two configurations: a random coil and an $\alpha $-helix. In the
coil state the monomers are free to rotate thus leading to a small Kuhn
length with typical value of $l_{c}\simeq 18\AA $ corresponding to the
length of few monomers. For simplicity we will identify the Kuhn length with
the span of a single monomer, $a\simeq 3.8\AA $. The helical configuration
involves hydrogen bonds between residues $i$ and $i+3$. This constrains the
orientation of the three intermediate monomers. As a result the persistence
length $P$ of the helical configuration is large, of order $P\simeq 2000\AA$.
At the same time the projected length of a monomer along the axis of the
helix decreases to $a_{h}\simeq 1.5\AA $. The length of the polypeptide in a
helical configuration is thus smaller than the span of a fully extended coil
by a factor of $\gamma =a_{h}/a\simeq 2/5$~\cite{Birshtein,Poland,Grosberg}.

The homopolypeptides are modeled as a two state system. Each monomer can
exist either in a helical or a coil state. Choosing the coil state as a
reference state, the excess free energy $\Delta f$ of a monomer within a
helical domain reflects two contributions. First is the change in enthalpy
upon formation of an intrachain $H$-bond, $\Delta h$. The second is
associated with the loss of configurational entropy, $\Delta s$, in the
helical state. Altogether
\begin{equation}
\Delta f \simeq \Delta h - T \Delta s.
\end{equation}
\noindent The enthalpy term favors the helical state while the entropy term
favors the coil state. Thus, at low temperatures the free energy $\Delta f$
is negative and the helical configuration is preferred while at high
temperatures $\Delta f$ is positive and the coil state is dominant. The two
terms are comparable at $T^{\ast }\simeq \Delta h/\Delta s$ when $\Delta
f\simeq 0$. $T^{\ast }$ provides an approximate value for the melting
temperature of the $\alpha $-helix.

An additional parameter is needed to describe the helix-coil transition. The
number of $H$-bonds in a helical domain consisting of $n$ monomers is $n-2$.
It is thus necessary to allow for the special state of the terminal monomers
of the helical domain. These two monomers lose their configurational entropy
with no gain due to the formation of $H$-bonds. Each of the terminal bonds
is thus assigned an excess free energy of
\begin{equation}
\Delta f_{t} \simeq - T \Delta s
\end{equation}
\noindent with respect to the coil state. In the following we will use 
$\Delta f_{t}$ measured with respect to the helical state, that is
\begin{equation}
\Delta f_{t} \simeq - \Delta h.
\end{equation}
\noindent As we shall discuss later, this definition is consistent with the
known experimental results. Clearly, this definition is consistent with the
traditional one when the coil state is used as a reference. $\Delta f_{t}$
plays the role of an interfacial free energy associated with the boundary
between helix and coil domains. This term makes the helix-coil transition
cooperative since it favors the formation of large domains.

It is customary to describe the helix-coil transition in terms of the
Zimm-Bragg parameters~\cite{Zimm}:
\begin{eqnarray}
s &=&\exp \left( -\Delta f/kT\right) , \\
\sigma  &=&\exp \left( -2\Delta f_{t}/kT\right) 
\end{eqnarray}
\noindent where $k$ is the Boltzmann constant. The $T$ dependent $s$
represents the Boltzmann factor associated with adding one monomer to a
helical domain. $s\simeq 1$ at the transition temperature $T^{\ast }$ while 
$s>1$ for $T<T^{\ast }$ and $s<1$ for $T>T^{\ast }$. $\sigma $ corresponds to
the Boltzmann factor associated with the creation of a helical domain. The
definition of $\Delta f_{t}$ as $-T\Delta s$ suggests that $\sigma $ is
independent of $T$. The definition of $\Delta f_{t}$ as $-\Delta h$ does not
predict that $\sigma $ is independent of $T$ but it yields an identical
estimate for the numerical value of $\sigma $~\cite{estimate}. The
experimentally measured $\sigma $, as obtained from plots of the helical
content {\em vs.} $T$, are in the range of $10^{-3}-10^{-4}$, depending on
the residue, and exhibit a weak $T$ dependence. As stated earlier, in the
following we will use $\sigma =\exp \left( 2\Delta h/kT\right) $. As we
shall discuss, a plot of the helical content {\em vs.} $s$ exhibits a
sigmoid behavior and the width of the transition scales with $T^{\ast
}\sigma ^{1/2}$. Thus, the transition becomes sharper, more cooperative, as 
$\sigma $ decreases.

\subsection{The Helix-Coil Transition: A Free Energy Approach}

Having defined the relevant molecular parameters, $\Delta f$ and $\Delta
f_{t}$ or equivalently $s$ and $\sigma$, we are in a position to analyze
the helix-coil transition as it occurs in free chains upon changing the
temperature $T$. The minimal description of the transition requires the
specification of two properties, the total number of monomers in a helical
configuration, $N\theta $ and the number of helical domains, $Ny$. While 
$\theta $ and $y$ are sufficient for a discussion of the helix-coil
transition of free chains, the analysis of the effect of extension requires
additional information. In particular, the probability distribution of
helical domains, $P_{h}(n)$, and of coil domains, $P_{c}(n)$, comprising of 
$n$ monomers. In order to determine these parameters we minimize the
appropriate phenomenological free energy. This method recovers the results
obtained by the transfer matrix approach and is easily generalized to allow
for the effect of applied tension.

The free energy allows for two principal contributions: (i) The free energy
of the monomers. All $N \theta$ helical monomers, including the terminal
monomers of the domains are assigned an excess free energy of $\Delta f$.
The $2Ny$ terminal monomers are assigned an additional free energy of 
$\Delta f_{t} = -\Delta h$. As a result their excess free energy with respect
to the coil state is $\Delta f_{t} = - T \Delta s$. (ii) A mixing entropy
term associated with the different possible placements of $N \theta$ helical
monomers and $2Ny$ terminal monomers constituting the domain boundaries.
Since the helical and coil domains alternate, there is no entropy due to
their mixing. Rather, the mixing entropy arises because of the
polydispersity of the domains {\em i.e.}, helical and coil domains of
different size $n$ are considered distinguishable. Since there are $Ny$
helical and $Ny$ coil domains the mixing entropy is
\begin{equation}  
\label{entropy2}
S_{mix} = - Nky \sum_{n=1}^{\infty} \left\{ P_h (n) \ln P_h (n) + P_c (n)
\ln P_c (n) \right\}.
\end{equation}
\noindent The free energy per monomer in the unperturbed chain is thus
\begin{eqnarray}
&&F_{0}/NkT = -\theta \ln s-y\ln \sigma   \label{eq_free2} \\
&&+y\sum_{n=1}^{\infty }P_{h}(n)\ln P_{h}(n)+y\sum_{n=1}^{\infty
}P_{c}(n)\ln P_{c}(n)  \nonumber \\
&&-\mu _{1}^{h}\left( \sum_{n=1}^{\infty }P_{h}(n)-1\right) -\mu
_{1}^{c}\left( \sum_{n=1}^{\infty }P_{c}(n)-1\right)   \nonumber \\
&&-\mu _{2}^{h}\left( \sum_{n=1}^{\infty }nP_{h}(n)-\frac{\theta }{y}
\right) - \mu_{2}^{c}\left( \sum_{n=1}^{\infty }nP_{c}(n)-\frac{1-\theta 
}{y} \right) .  \nonumber
\end{eqnarray}
\noindent The Lagrange multipliers, $\mu _{1}^{h}$ and $\mu _{1}^{c}$,
ensure the normalization $\sum_{n=1}^{\infty }P_{h}(n)=1$ and 
$\sum_{n=1}^{\infty }P_{c}(n)=1$. The two remaining Lagrange multipliers, 
$\mu_{2}^{h}$ and $\mu_{2}^{c}$, impose the average sizes $k_{h}=\theta/y$
and $k_{c}=(1-\theta )/y$ of the helical and coil domains. This description
corresponds to a simplified version of the transfer matrix method involving
a two by two matrix. In both cases there is no constraint ensuring that
helical domains incorporate at least three monomers~\cite{constraint}. Since
we focus on the case of $\sigma \ll 1$ and $N\gg 1$ the weight of small
domains is negligible and this description yields the correct results.

The probabilities $P_{h}(n)$ and $P_{c}(n)$ are determined by minimization
of $F_{0}$ with respect to $P_{h}(n)$ and $P_{c}(n)$ (Appendix A) leading to

\begin{eqnarray}
P_{h}(n) & = & \frac{y}{\theta - y} \left( \frac{\theta - y}{\theta}
\right)^{n},  \label{proba_h} \\
P_{c}(n) & = & \frac{y}{1-\theta - y} \left( \frac{1-\theta - y}{1- \theta}
\right)^{n}.  \label{proba_c}
\end{eqnarray}
\noindent Note that $P_{h}(n)$ and $P_{c}(n)$ decay exponentially with 
decay constants $\lambda _{h}=-1/\ln (1-y/\theta)$ and $\lambda_{c}
=-1/\ln \left[1-y/(1-\theta)\right]$. Thus, $P_{h}(n)$ and $P_{c}(n)$ 
are not sharply peaked at the mean values $k_{h}=\theta/y$ and 
$k_{c}=(1-\theta)/y$. In particular, there is a significant population 
of small helices even when the average size of the helical domain 
$k_{h}=\theta/y$ is large. In this limit, $\lambda_{h}\simeq k_{h}$. 
Similarly, when $k_{c}=(1-\theta)/y$ is large $\lambda_{c}\simeq k_{c}$.

Upon substituting $P_{h}(n)$ and $P_{c}(n)$ in $F_{0},$ as given by 
(\ref{eq_free2}), we obtain $F_{0}$ as a function of $\theta $ and $y$ 
only~\cite{smix}:

\begin{eqnarray}  
\label{eq_free1}
\frac{F_{0}}{NkT} & = & -\theta \ln s - y \ln \sigma + (\theta - y) \ln 
\frac{\theta-y}{\theta} + y \ln \frac{y}{\theta} \\
&& + (1-\theta-y) \ln \frac{1-\theta-y}{1-\theta} + y \ln 
\frac{y}{1-\theta}. \nonumber
\end{eqnarray}

\noindent The equilibrium conditions $\partial F_{0}/\partial \theta = 0$
and $\partial F_{0}/\partial y = 0$ then lead to

\begin{eqnarray}
(1-\theta)(\theta-y) & = & s \, \theta (1-\theta-y), \\
y^{2} & = & \sigma \, (\theta-y)(1-\theta-y).
\end{eqnarray}

\noindent In turn, these yield (Appendix B) the expression for the
equilibrium values of $\theta$ and $y$

\begin{eqnarray}
\theta &=&\frac{1}{2}+\frac{1}{2}\frac{s-1}{\sqrt{(s-1)^{2}+
4\,\sigma s}}, \\
y &=&\frac{2\,\sigma s\left[ (s-1)^{2}+4\,\sigma s\right]^{-1/2}}{s+1+
\sqrt{(s-1)^{2}+4\,\sigma s}}.
\end{eqnarray}

\noindent These expressions are identical to the ones obtained {\em via} the
transfer matrix method~\cite{Grosberg}.

\begin{figure}[tbp]
\centerline{\epsfig{file=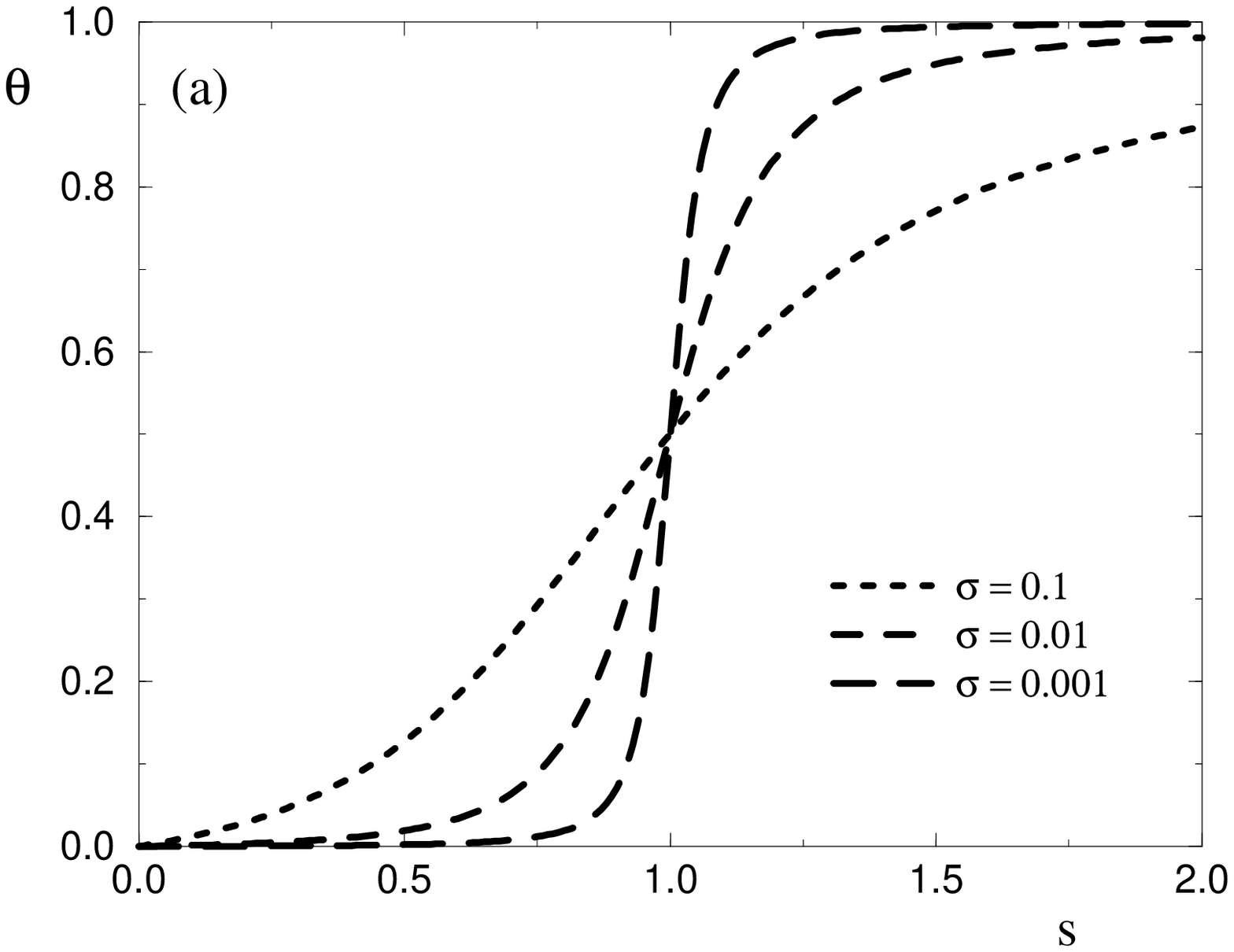,width=3in}}
\centerline{\epsfig{file=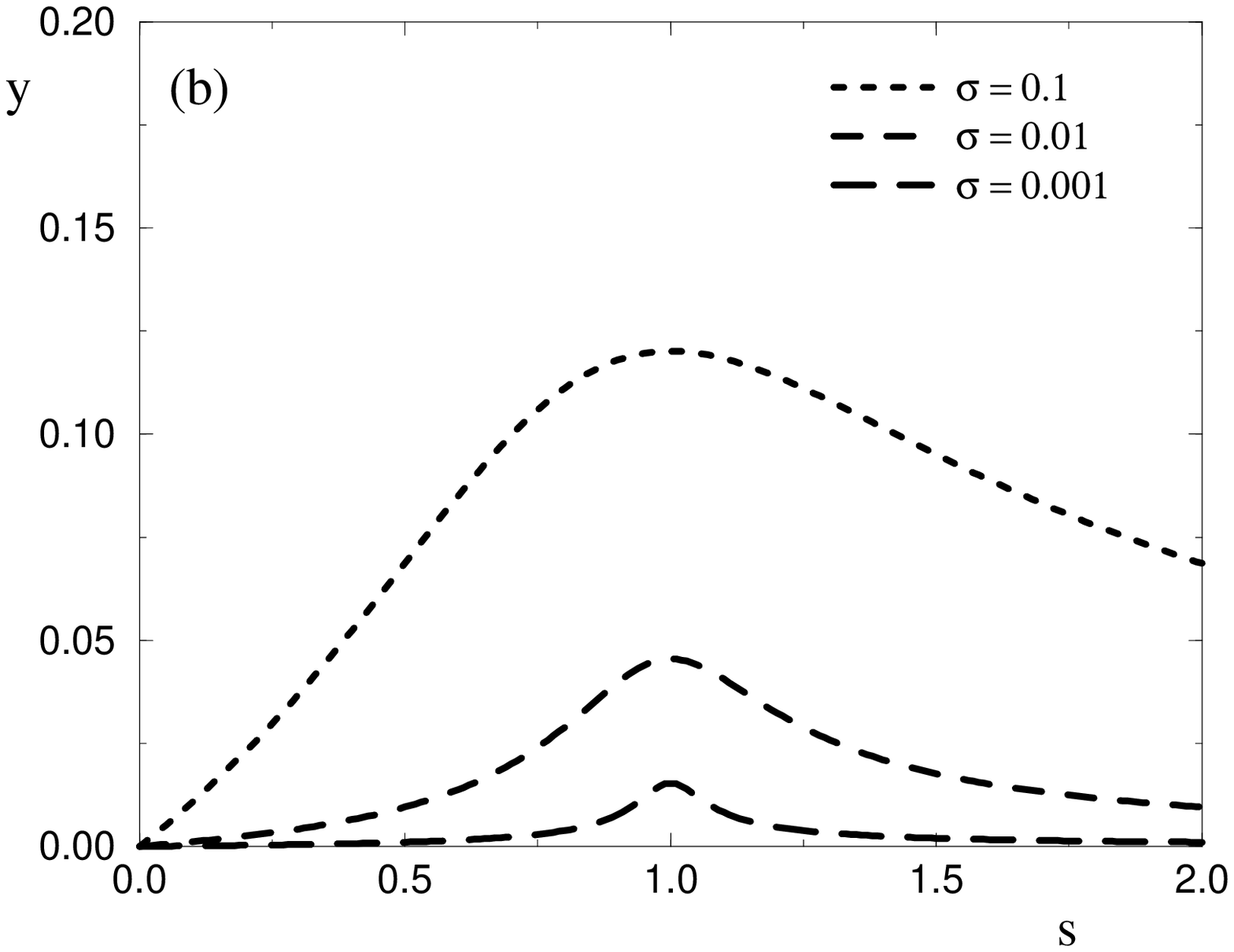,width=3in}}
\caption{\label{theta0} The helical fraction $\theta$ (a) and the domain 
fraction $y$ (b) are plotted as a function of $s$ for long polypeptides 
free of tension. In all cases, the helix-coil transition occurs in the 
vicinity of $s=1$ however, the width of the transition region decreases 
with $\sigma$.}
\end{figure}

The behavior of $\theta $ and $y$ as a function of $s$ and $\sigma $ is
plotted in Fig.~\ref{theta0}. 
When $\sigma =1$ we recover the Boltzmann distribution for
independent two-level systems: $\theta =s/(1+s)$. In marked contrast when 
$\sigma =0$, the fraction of helical monomers is steplike: $\theta =1$ for 
$s>1$ and $\theta =0$ for $s<1$. The chain is either a pure helix or a pure
coil. For intermediate values of $\sigma $, the plot of $\theta $ {\em vs.} 
$s$ yields a sigmoid curve with a transition region of width $\sigma ^{1/2}$
centered around $s=1$. The number of domains is maximal in the vicinity of 
$s\simeq 1$ and it decreases with $\sigma $. The parameter $s=\exp \left(
-\Delta f/kT\right) $ serves as a measure of the temperature, $T$. For low
temperatures, $s\gg 1$ and the polymer is mainly in a helical state. At high
temperatures, $s\ll 1$ and the polymer is mostly in a coil state. The
crossover occurs at $T^{\ast }=\Delta h/\Delta s$ for which $s=1$. Near this
crossover, $s-1\sim T^{\ast }-T$ and the width of the crossover regime is 
$\Delta T\sim T^{\ast }\sigma ^{1/2}$.

\section{Extension Behavior: Qualitative Analysis}

The most notable feature of the force laws of homopolypeptides are the
plateaus. In this section we present qualitative and semiquantitative
analysis of the plateaus. Each plateau in the force law corresponds to a
coexistence between two states, a helix and a coil. The situation is
somewhat analogous to a liquid-gas transition which leads to a plateau in
the pressure-volume diagram. In our case, the tension $f$ plays the role of
the pressure and the end-to-end distance $R$ is analogous to the volume.
This analogy is however of limited validity since the homopolypeptides are
one-dimensional systems with short range interactions and thus incapable of
undergoing a first order phase transition. In such systems the boundary free
energy is independent of the size of the domain. As a result, a first order
phase transition involving a coexistence of two domains is prevented by the
mixing entropy~\cite{Landau}. Accordingly, the average domain sizes 
$k_{c}=(1-\theta )/y$ and $k_{h}=\theta /y$, as obtained in the previous
section, are independent of the size of the chain, $N$. Nevertheless, the
main features of the system can be recovered if one ignores the role of the
mixing entropy. As expected, within this rough approximation the plateaus
are erroneously associated with first order phase transitions. In spite of
this deficiency, this approximation recovers the correct length and force
scales but the force laws obtained are wrong in detail. Before we present
this approximation we investigate the crossing of the free energy curves
associated with the ``pure'' helix and coil states. This is a rough
diagnostic for first order phase transitions. In our case it provides an
insight concerning the physical origin of the plateaus as well as a
reasonable estimate for some of the length and force scales involved.

\subsection{Free Energy Curve Crossing}

The occurrence of the plateaus is signaled by the crossing of the free
energy curves of the pure helix and the pure coil as a function of $R$. Each
crossing point signals a coexistence between the two ``phases'' and thus a
plateau in the force law.

With the unperturbed coil as a reference state, the free energy of the coil
is simply its elastic free energy. In this section, we will utilize the
fixed $R$ ensemble and we thus denote the elastic free energy by $F_{el}(R)$.
To allow for the finite extensibility of the chain we use the freely jointed
chain (FJC) model (Appendix C) for $F_{el}(R)$ and the free energy per
monomer is  
\begin{equation}
\frac{F_{el}(R)}{NkT}=rx-{\cal L}_{int}(x)
\end{equation}
\noindent where ${\cal L}_{int}(x)=\ln \left[ \sinh (x)/x\right] $, $r=R/Na$
is the reduced end-to-end distance of the polymer and $x=fa/kT$ is the
reduced force. $r$ and $x$ are related {\it via} $r={\cal L}(x)\equiv
\coth (x)-1/x$ where ${\cal L}(x)$ is the Langevin function.

The free energy of the helix reflects two contributions. First is the excess
free energy of a helix, $F_{h}=N\Delta f+2\Delta f_{t}$. In the limit of an
infinite chain, when the contribution of the terminal monomers is
negligible, the free energy per monomer is $F_{h}/NkT=-\ln s$. At this stage
we assume that the persistence length of the helix is infinite and that it
may be viewed as a perfectly rigid and unextendable rod. Within this view,
extending the helix beyond its natural length is associated with an infinite
free energy penalty. This penalty is important because the helix is shorter
than the fully extended coil. As noted earlier, the contribution of each
helical monomer to the overall length of the helix is $a_{h}=\gamma a<a$.
Altogether, $F_{h}/NkT=-\ln s$ so long as $r=R/Na\leq \gamma $ while in
the range $r>\gamma $ it is $F_{h}/NkT=\infty$.

\begin{figure}[tbp]
\begin{center}
\epsfig{file=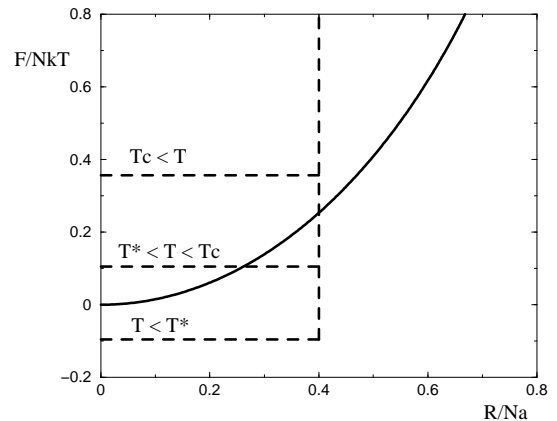,width=3in}
\caption{\label{crossing}
The occurrence of plateaus in the extension force laws is signaled
by crossing of the free energy curves corresponding to a polypeptide in a
pure coil form (continuous curve) and in a fully helical form (dashed
lines). }
\end{center}
\end{figure}

Three scenarios are now possible (Fig.~\ref{crossing}). At low temperatures, 
$T<T^{\ast}$ and $s>1$. Accordingly, $F_{h}/NkT<0$ while $r\leq \gamma$. In
this regime the helix is more stable than the coil. When $r>\gamma $ the
free energy of the helix diverges and the coil state becomes preferable.
This curve crossing corresponds to a plateau involving the breakdown of the
helix due to the applied tension. Above the melting temperature, for 
$T\gtrsim T^{\ast }$ and $F_{h}/NkT = -\ln s \gtrsim 0$ 
the unstretched polypeptide is in a coil state. 
Upon stretching $F_{el}/NkT$ increases and eventually
intersects with $F_{h}/NkT$. This intersection signals a plateau in the
force law due to the formation of helical domains. At $r=\gamma $ the free
energies $F_{el}/NkT$ and $F_{h}/NkT$ cross again because of the infinite
free energy of a helix with $r>\gamma $. This second intersection signals a
second plateau involving the melting of the helix and the formation of coil
domains. Finally, for sufficiently high temperatures, $T>T_{c}$, no
intersections occur and the polypeptide remains in a coil state for all $r$
values. The threshold temperature for this regime, $T_{c}$, is defined by
the equality $F_{el}(N\gamma a)=F_{h}$ where $F_{el}(N\gamma a)$ is the free
energy of a stretched coil with $r=\gamma $.

\subsection{The $S_{mix}=0$ approximation}

The intersection of the free energy curves provides a rough indicator for
the occurrence of plateaus in the force law. However, it does not allow to
calculate an explicit force-extension diagram. A simple approach yielding
the main features of the force law is the $S_{mix}=0$ or diblock
approximation~\cite{Buhot,BH}. Within this approximation the force law is
calculated for a diblock copolymer with one helical domain and one coil
domain. The size of the annealed domains varies with the extension and is
determined by the equilibrium conditions. As is implied by its name, in this
approximation $S_{mix}$ is neglected. As expected, setting $S_{mix}=0$ leads
to a first order phase transition associated with a coexistence of two
``phases''. Consequently, some features of the force law are wrong. In
particular, this approximation yields a perfect plateau with 
$f(R)=const^{\prime }$ while in reality $f$ increases weakly with $R$. These
features notwithstanding, this simple approximation does recover the correct
length and force scales. Within this discussion we will also briefly examine
the effect of the persistence length of the helical domains. In particular
we compare the behavior of a helix endowed with an infinite persistence
length to that of a helix with a finite one.

Within the $S_{mix}=0$ or diblock approximation the free energy $F_{chain}$
of the chain is

\begin{equation}
F_{chain} = N \theta \Delta f + 2 \Delta f_{t} + F_{el}(\theta ,\gamma ,R).
\end{equation}

\noindent Here, the first term allows for the excess free energy of the
helical monomers. Since $S_{mix}=0$ the helical and coil phases separate
into two domains. The ``interfacial'' free energy of the single helical
domain gives rise to the second, constant term. These two terms are
supplemented by an elastic free energy $F_{el}$ that depends on $\theta $, 
$\gamma $ and the end-to-end distance, $R$. Different versions of the 
$S_{mix}=0$ are possible, depending on the choice of $F_{el}$. In the
following we focus on the simplest version where $F_{el}$ is given by the
Gaussian elasticity of an ideal random coil in the fixed $R$ ensemble. 
Two additional assumptions are invoked: (i) The helical domain is perfectly
rigid and unextendable and (ii) the helical domain is perfectly aligned with
the applied force. These assumptions specify the elastic free energy per
monomer 
\begin{equation}
\frac{F_{el}}{NkT}=\frac{3}{2}\frac{(r-\gamma \theta )^{2}}{1-\theta }.
\end{equation}
\noindent This simple version of the diblock approximation permits an
explicit analytical solution for the force law. The corresponding
equilibrium condition is 
\begin{equation}
\frac{\partial F_{chain}}{\partial \theta }=0=\frac{3}{2}
\frac{(r-\gamma \theta )^{2}}{(1-\theta )^{2}}-3\gamma \frac{r-\gamma 
\theta }{1-\theta }+\frac{\Delta f}{kT}.
\end{equation}
\noindent Within this approximation the equilibrium condition $\partial
F_{chain}/\partial y=0$ is an identity yielding no additional information.
The solution of this quadratic equation in $X=(r-\gamma \theta)/(1-\theta)$
is $X_{\pm }=\gamma \left( 1\pm \sqrt{1-\frac{2\,\Delta f}{3\gamma^{2}kT}}
\right)$. This defines a critical $\Delta f$ 
\begin{equation}
\Delta f_{c}/kT_c=3\gamma ^{2}/2
\end{equation}
\noindent corresponding to a critical $s_{c}=\exp (-\Delta f_{c}/kT_c)$ or 
$T_{c}$. While $\Delta f>\Delta f_{c}$, the equilibrium condition can not be
satisfied, that is, the polypeptide remains in a coil state irrespective of
the imposed tension. A critical point occurs for $\Delta f=\Delta f_{c}$.
When $0<\Delta f<\Delta f_{c}$ two solutions exist, each defining a plateau
in the force {\em vs.} extension curve. The solution $X_{-}$ corresponds to
the lower plateau, obtained for $r<\gamma $, where 
\begin{equation}
\theta =1-\frac{1-r/\gamma }{\sqrt{1-\Delta f/\Delta f_{c}}}.
\end{equation}
\noindent This first regime disappears for $\Delta f<0$ or 
$T<T^{\ast }$. The condition $\theta =0$ defines a lower boundary of
the plateau, 
\begin{equation}
r_{1L}=\gamma \left( 1-\sqrt{1-\Delta f/\Delta f_{c}}\right) .
\end{equation}
\noindent $r_{1L}$ increases as $\Delta f$ increases until $r_{1L}=\gamma $
for $\Delta f=\Delta f_{c}$. For $r<r_{1L}$ the fraction of helices is 
$\theta =0$ and the chain is in a purely coil form. The upper boundary of
this plateau occurs at 
\begin{equation}
r_{1U}=\gamma 
\end{equation}
\noindent when the chain is fully helical, $\theta =1$. At intermediate
extensions, $r_{1L}\leq r\leq r_{1U}$ the fraction of helical bonds
increases as 
\begin{equation}
\theta =\frac{r-r_{1L}}{r_{1U}-r_{1L}}
\end{equation}
\noindent and $\theta =1/2$ occurs at the midpoint of the plateau 
$r_{1}(\theta =1/2)=(r_{1U}+r_{1L})/2$. The equilibrium free energy of the
chain for $r<r_{1L}$ is $F_{chain}/NkT=3r^{2}/2$ while for $r_{1L}\leq r\leq
r_{1U}$ it is 
\begin{equation}
\frac{F_{chain}}{NkT}=\frac{3r_{1L}^{2}}{2}\frac{r_{1U}-r}{r_{1U}-r_{1L}}+
\frac{\Delta f}{kT}\frac{r-r_{1L}}{r_{1U}-r_{1L}}.
\end{equation}
\noindent The second plateau, is described by the solution, $X_{+}$, leading
to 
\begin{equation}
\theta =1-\frac{r/\gamma -1}{\sqrt{1-\Delta f/\Delta f_{c}}}.
\end{equation}
\noindent The lower boundary of this plateau occurs at 
\begin{equation}
r_{2L}=r_{1U}=\gamma 
\end{equation}
\noindent when $\theta =1$. The upper boundary, when $\theta =0$, occurs at 
\begin{equation}
r_{2U}=\gamma \left( 1+\sqrt{1-\Delta f/\Delta f_{c}}\right) .
\end{equation}
\noindent For $r_{2L}\leq r\leq r_{2U}$ the fraction of helical monomers
decreases as 
\begin{equation}
\theta =\frac{r_{2U}-r}{r_{2U}-r_{2L}}
\end{equation}
\noindent and $\theta =1/2$ occurs at the midpoint of the plateau 
$r_{2}(\theta =1/2)=(r_{2U}+r_{2L})/2$. Notice that the width of the two
plateaus shrinks as $\Delta f\rightarrow \Delta f_{c}$ or $T\rightarrow 
T_{c}$. For $r\geq r_{2U}$ the fraction of monomers in a helical state is 
again $\theta =0$ and the corresponding free energy is 
$F_{chain}/NkT=3r^{2}/2$.
The equilibrium free energy of the chain in the range $r_{2L}\leq r\leq
r_{2U}$ is 
\begin{equation}
\frac{F_{chain}}{NkT}=\frac{3r_{2U}^{2}}{2}\frac{r-r_{2L}}{r_{2U}-r_{2L}}+
\frac{\Delta f}{kT}\frac{r_{2U}-r}{r_{2U}-r_{2L}}.
\end{equation}
\noindent Note that only the lower plateau is obtained if $\gamma =1$ is
assumed\cite{Buhot}.

\begin{figure}[tbp]
\begin{center}
\epsfig{file=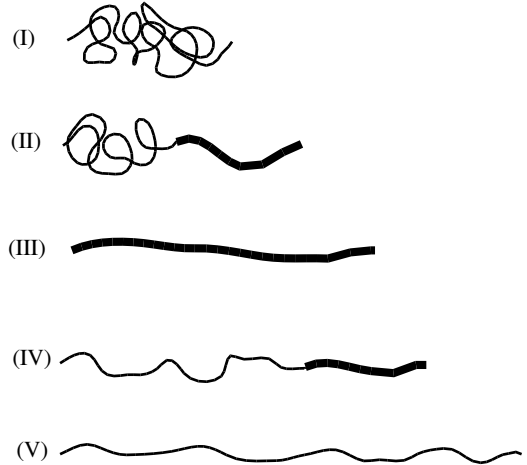,width=3in}
\caption{\label{phases}
Schematic pictures of the stretching of a polypeptide just above
$T^{\ast}$ within the $S_{mix}=0$ approximation: (I) for small extension
the chain is a weakly extended coil, (II) for $r > r_{1L}$ a helical domain
coexists with a weakly extended coil domain, (III) at $r = \gamma$
the chain is fully helical, (IV) a helical domain coexists with a
strongly extended coil domain for $r>\gamma$ and eventually, (V) for $r >
r_{2U}$ the chain is again fully in a coil state but in a highly
stretched configuration.}
\end{center}
\end{figure}

The $S_{mix}=0$ approximation allowed us to obtain rough force laws
corresponding to the three scenarios identified by the curve crossing
argument. Thus, for $\Delta f>\Delta f_{c}$ or $T>T_{c}$, the polypeptide
remains in a coil configuration for the whole range of extension. The ``two
plateau scenario'' occurs for $0<\Delta f<\Delta f_{c}$ corresponding to 
$s_{c}<s<1$ or $T^{\ast }<T<T_{c}$. In this case we can distinguish five
regimes (Fig.~\ref{phases}) : 
(I) For low extensions the chain is in a coil state, 
$\theta =0$, and the elastic behavior is Gaussian leading to $fa/kT=3r$. 
(II) In the range $r_{1L}\leq r\leq r_{1U}$ helix and coil domains coexist
according to the lever rule $\theta /(1-\theta )=(r-r_{1L})/(r_{1U}-r)$.
This coexistence is associated with a plateau with a constant tension 
\begin{equation}
\frac{f_{1co}a}{kT}=3r_{1L}.
\end{equation}
\noindent (III) At $r_{1U}=r_{2L}=\gamma $ the force curve exhibits a
discontinuity associated with a step like increase to a second plateau. The
magnitude of the step, $\delta f$ is 
\begin{equation}
\delta fa/kT=6\gamma \sqrt{1-\Delta f/\Delta f_{c}}.
\end{equation}
(IV) In the range $r_{2L}\leq r\leq r_{2U}$ helix and coil domains coexist
according to the lever rule $\theta /(1-\theta )=(r_{2U}-r)/(r-r_{2L})$.
This coexistence is associated with a second plateau with a constant tension 
\begin{equation}
\frac{f_{2co}a}{kT}=3r_{2U}.
\end{equation}
\noindent (V) Finally, for $r>r_{2U}$ the chain is in a strongly
stretched coil state obeying $fa/kT=3r$.

When $\Delta f<0,$ or $T<T^{\ast },$ the unperturbed chain is in a helical
state. Consequently, the lower plateau disappears. It is replaced by a
``quasi plateau'' which is not associated with a helix-coil coexistence.
Rather, it is due to the facile orientation of the helical chain which is
modeled as a rigid rod. At $r=\gamma $ a jump occurs to the upper plateau. 

The analysis presented above utilized the Gaussian elastic free energy in
the fixed $R$ ensemble to describe the coil. While this allows us to obtain
explicit expressions for all the characteristics of the force curve, it also
introduces errors. Clearly, this choice of $F_{el}$ does not capture the
behavior of the coil in the high extension regime, where finite
extensibility begins to play a role. As a result, the values of $r_{2U}$ 
are overestimated. This has two consequences. First, within this
approximation the span of the two plateaus are equal $r_{1U}-r_{1L}=
r_{2U}-r_{2L}=\gamma \sqrt{1-\Delta f/\Delta f_{c}}$ while for
more realistic choice of $F_{el}$ the higher plateau is narrower. A second
more important pathology is the disappearance of regime (V) in the $\Delta
f<0$ case.

\begin{figure}[tbp]
\begin{center}
\epsfig{file=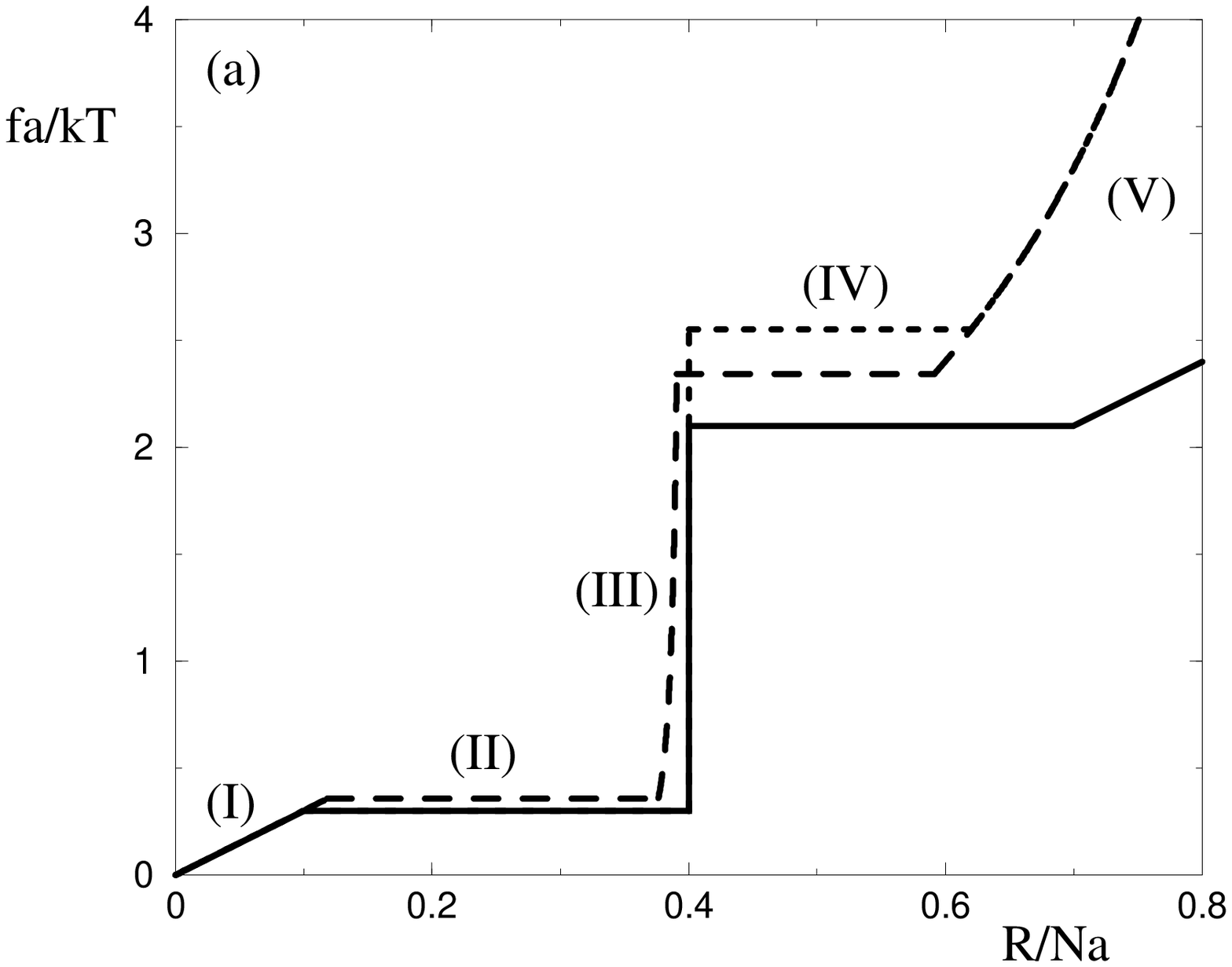,width=3in}
\epsfig{file=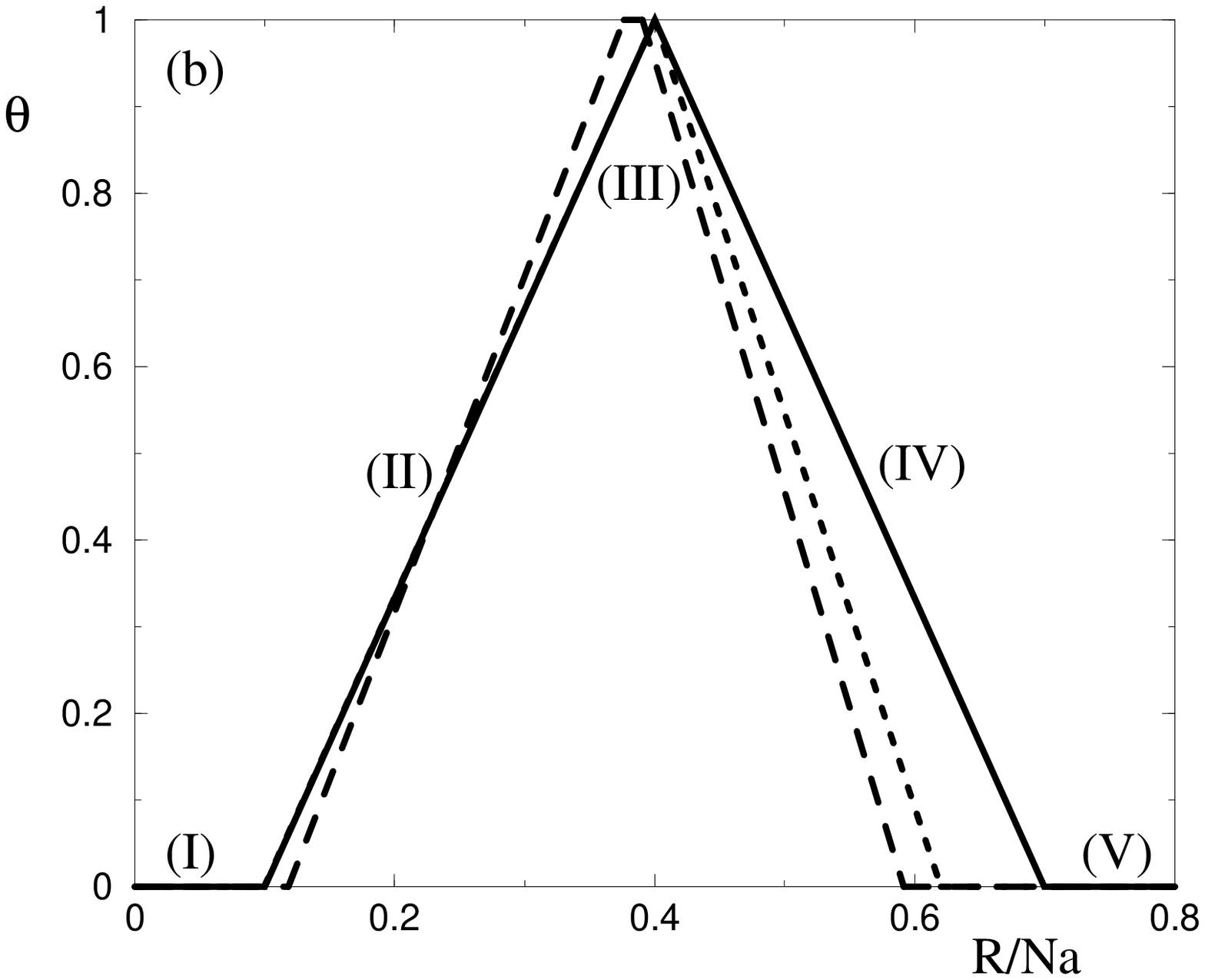,width=3in}
\caption{\label{force_db}
Plots of the reduced force $x=fa/kT$ (a) and the helical fraction 
$\theta$ (b) {\em vs.} the reduced end-to-end distance, $r=R/Na$ as
obtained via the $S_{mix}=0$ approximation for the $s_{c} < s < 1$ scenario.
Both figures correspond to $s=0.9$, $\gamma = 2/5$ and $n_h = 1000$.
The continuous curve corresponds to the case of rod like helices and
Gaussian coils. The dashed lines depict the results obtained when the coils
are described by the FJC model and the helices as rigid rods (short dashed
line) or by the WLC model (long dashed line). The roman numerals identify
the different regimes as depicted in Fig.~\ref{phases}.}
\end{center}
\end{figure}

\begin{figure}[tbp]
\begin{center}
\epsfig{file=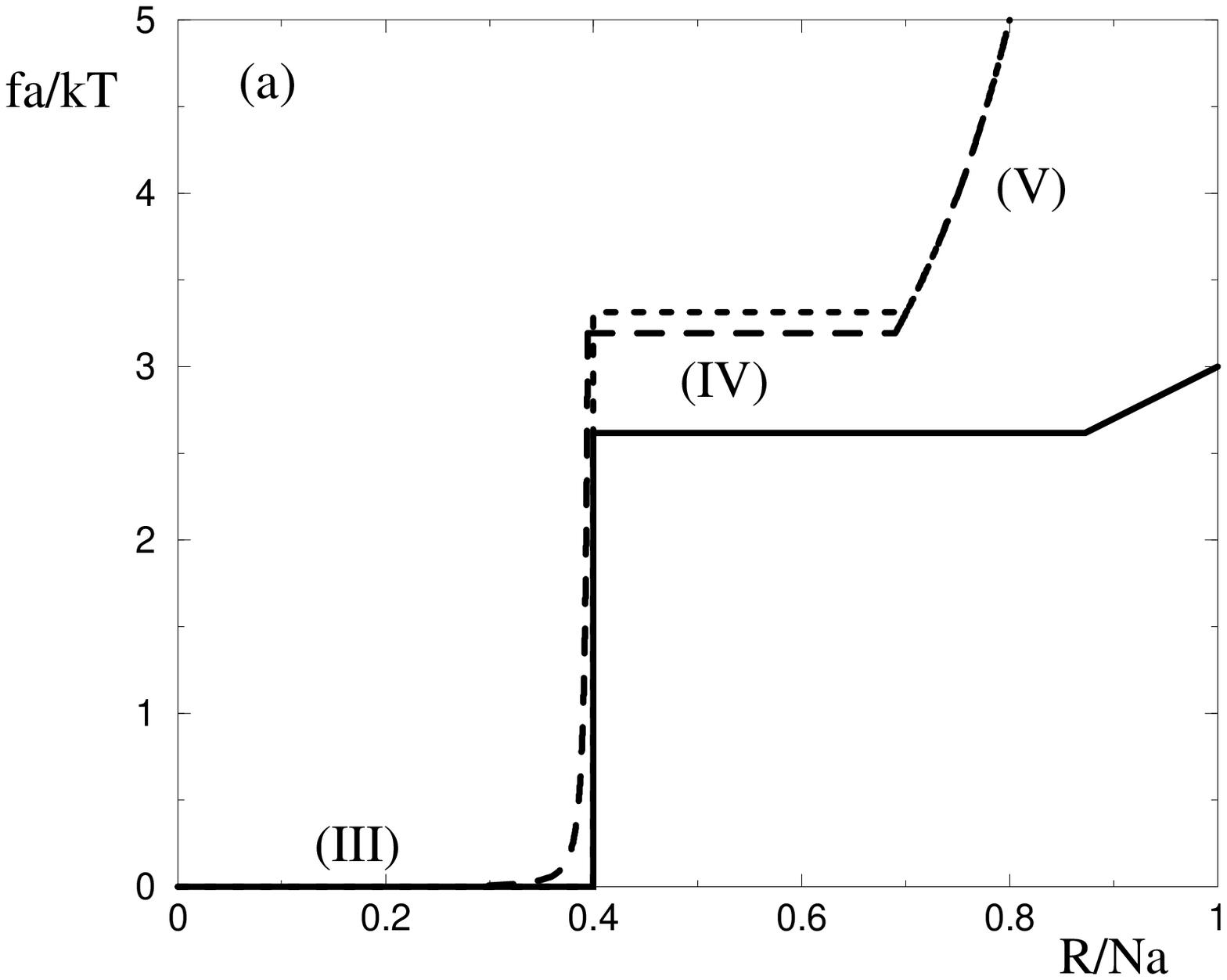,width=3in}
\epsfig{file=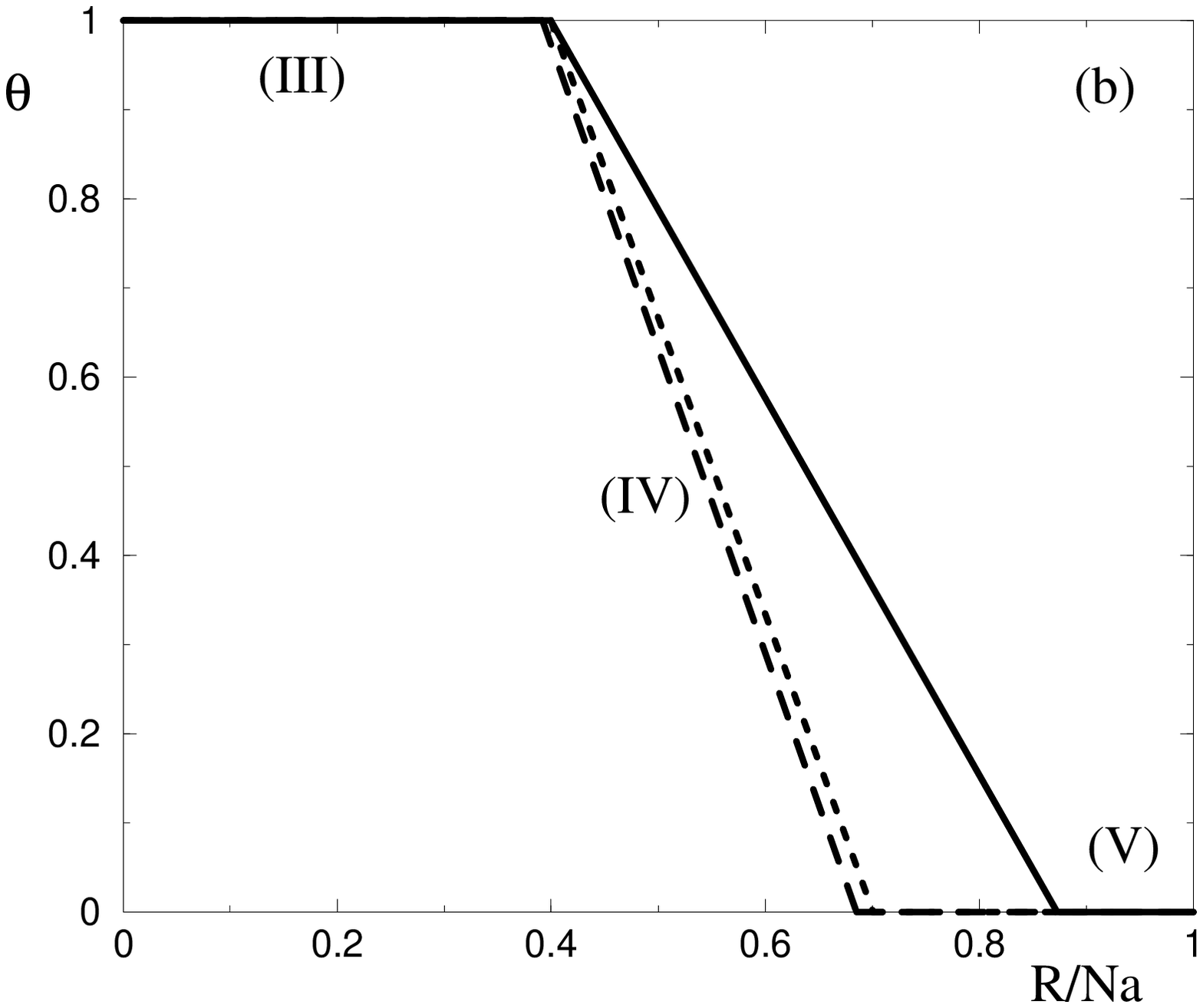,width=3in}
\caption{\label{force_db2}
Same as previous figures for the $s > 1$ scenario. Both figures
correspond to $s=1.1$, $\gamma = 2/5$ and $n_h = 1000$.}
\end{center}
\end{figure}

To allow for the finite extensibility it is necessary to use the freely
jointed chain model (FJC) for the coil segment~\cite{footnote2}. It is
important to note that so long as $\gamma <0.5$ and $\Delta f>0$ the
discrepancies between the two schemes are rather small 
(Fig.~\ref{force_db}). Another
ingredient of our simplified analysis is the modeling of the helices as
rigid and unextendable rods. An alternative description of the elastic
properties of helices models them as worm like chains {\em i.e.,}
semiflexible chains capable of undergoing undulations 
(see Appendix D)~\cite{footnote3}. 
As we shall see later, this semiflexible
approximation of the helices has a qualitative effect on the results of the
rigorous analysis of the system. However, in the context of the $S_{mix}=0$
approximation for $\Delta f>0$, it mainly affects the transition between the
two plateaus (Fig.~\ref{force_db}). 
When the helix is modeled as a rigid rod the
transition occurs as a jump that takes place at $r=\gamma $. On the other
hand, when the WLC model is used the jump, while abrupt, takes place over a
finite interval {\em i.e.,} $r_{1U}=r_{2L}$ is replaced by 
$r_{1U}<r_{2L}<\gamma$. For the $\Delta f<0$ (Fig.~\ref{force_db2}) 
the transition at the
upper boundary of the ``quasi plateau'' is more gradual when the elasticity
of the semiflexible helix is allowed for. Finally, note that the diblock
approximation is recovered from the more rigorous treatments presented below
when the $N\rightarrow \infty $ limit is taken before the $\sigma
\rightarrow 0$ limit. When $\sigma \rightarrow 0$ limit is approached for a
finite chain it leads to a single domain.

\section{Rigorous analysis}

The analysis presented in the previous section fails to allow for a number
of important ingredients. The most important is the mixing free energy, 
$-TS_{mix}$, associated with the polydispersity of the domains. In this
section we present an analysis that allows for the role of this mixing
entropy. Because of $S_{mix}$ the helix-coil transitions no longer take
place as first order phase transitions. In turn this has two important
consequences: (i) The ``perfect'' plateaus, with constant forces $f_{1co}$
and $f_{2co}$ are replaced by ``imperfect'' plateaus with a finite slope.
(ii) The crossovers between the plateaus and the neighboring regimes become
smoother.

The analysis is based on a generalization of the approach described in
section II that is, all chain characteristics, $P_{h}(n)$, $P_{c}(n)$, 
$\theta$ and $y$ are obtained from the minimization of the appropriate free
energy {\em i.e.}, the equilibrium conditions. As in section III, the free
energy is of the form

\begin{equation}
F_{chain}=F_{0}+F_{el}
\end{equation}
\noindent where, $F_{0}$ is the free energy of the unperturbed free chain
and $F_{el}$, accounts for the elasticity of the chain. However, in the
following we utilize different expressions for these two terms. Since we aim
to obtain $P_{h}(n)$ and $P_{c}(n)$ it is necessary to utilize $F_{0}$ as
given by equation (\ref{eq_free2}). The handling of the elastic free energy
is significantly different. In section III we used the Gaussian elasticity
to describe the deformation behavior of the coils and assumed that the
helices contribute to the elastic behavior as a geometric constraint. While
this rough description is sufficient at the level of the $S_{mix}=0$
approximation, a more refined treatment is needed for the detailed analysis
to be undertaken below. In particular, it is necessary to allow for two
additional features: (i) The finite extensibility of the chain and (ii) the
gradual alignment of the helical domains with the applied tension. The
freely jointed chain (FJC) model allows us to incorporate the finite
extensibility of the coil domains. With regard to the helices we will
explore two different descriptions that impose finite extensibility on the
helices. First we will model the helical domains as rigid and unextendable
rods and allow for their length dependent alignment by using the FJC model 
{\em i.e.,} the helical domains are considered as freely jointed rigid rods
forming a chain. This amounts to the assumption of an infinite persistence
length for each helical domain. This picture is reasonable so long as the
size of the helical domains is small compared to the persistence length, 
$P$. In the second approach, each of the helical domains is viewed as worm 
like chain (WLC) that is, semiflexible chain characterized by a finite
persistence length and capable of undergoing undulations. It is assumed that
the domains are large in comparison to $P$. An important difference between
the FJC and the WLC models is that in the later the persistence length
depends on the applied tension. Our approach utilizes the analysis of Marko
and Siggia of the extension elasticity of the WLC~\cite{Marko}. The
utilization of these models makes the constant $f$ ensemble mathematically
convenient. This choice of ensemble is also preferable because it
corresponds better to the experimental situation encountered in single
molecule measurements. In each case, we first obtain the equilibrium values
of $P_{h}(n)$, $P_{c}(n)$, $\theta $ and $y$ of a chain subject to a given
tension, $f$, by minimizing $F_{chain}(f)$ with respect to $P_{h}(n)$, 
$P_{c}(n)$, $\theta $ and $y$. This allows us to determine the equilibrium 
$F_{chain}(f)$ thus enabling us to calculate the $fR$ diagram by using 
\begin{equation}
R=-\partial F_{chain}(f)/\partial f.  
\label{Rdef}
\end{equation}

\subsection{Rigid helices}

When the helical domains are viewed as rigid rods and the FJC is invoked,
the reduced end-to-end distance $r=R/Na$ of the polypeptide is 
\begin{equation}
\label{r_rigid}
r=y\sum_{n=1}^{\infty }P_{c}(n)\,n\,{\cal L}(x)+y\sum_{n=1}^{\infty
}P_{h}(n)\,n\gamma \,{\cal L}(n\gamma x)
\end{equation}

\noindent where $x=fa/kT$ is the reduced tension. The first part of this
expression corresponds to the projection of the coil monomers along the
direction of the tension. The second part reflects to the contribution of
the helical domains to $r$. Because the individual monomers within the coil
domains are aligned independently, the domain size does not play a role.
The extension of each coil domain is extensive with respect to the number 
of monomers, $n$. Consequently, this contribution may be expressed in terms 
of the first moment of the probability distribution $k_{c}=(1-\theta
)/y=\sum_{n}nP_{c}(n)$. As a result the full probability $P_{c}(n)$ is
irrelevant. In marked contrast, the monomers within the helical domains are
not aligned individually but as part of the rod-like domains. The
non-linearity introduced by the ${\cal L}(n\gamma x)$ term in the expression
for $r$ couples $f$ and $P_{h}(n)$. As a consequence, $P_{h}(n)$ will be
explicitly modified by the applied tension while $P_{c}(n)$ depends on $f$
only implicitly, {\it via} $\theta $ and $y$.

Eliminating $P_{c}(n)$ from (\ref{r_rigid}), we obtain 
\begin{equation}
r=(1-\theta ){\cal L}(x)+y\sum_{n=1}^{\infty }P_{h}(n)\,n\gamma 
\,{\cal L} (n\gamma x).  \label{Rdef2}
\end{equation}
\noindent This expression for $R(f)$ enables us to obtain the
corresponding elastic free energy $F_{el}(f)=-\int_{0}^{f}Rdf^{\prime }$
(Appendix C) 
\begin{equation}
\frac{F_{el}}{NkT}=-(1-\theta ){\cal L}_{int}(x)-y\sum_{n=1}^{\infty }
P_{h}(n)\, {\cal L}_{int}(n\gamma x)
\end{equation}
\noindent where ${\cal L}_{int}(x)=\ln \left[ \sinh (x)/x\right] $.

Altogether the free energy per monomer is thus 
\begin{eqnarray}
&& F_{chain}/NkT =-\theta \ln s-y\ln \sigma  
\label{eq_free_rigid} \\
&&-(1-\theta ){\cal L}_{int}(x)
-y\sum_{n=1}^{\infty }P_{h}(n)\,{\cal L}_{int}(n\gamma x) \nonumber \\
&& +y\sum_{n=1}^{\infty }P_{h}(n)\ln P_{h}(n)+y\sum_{n=1}^{\infty
}P_{c}(n)\ln P_{c}(n) \nonumber \\
&& -\mu _{1}^{h}\left( \sum_{n=1}^{\infty }P_{h}(n)-1\right) -\mu
_{1}^{c}\left( \sum_{n=1}^{\infty }P_{c}(n)-1\right)  \nonumber \\
&& -\mu_{2}^{h}\left( \sum_{n=1}^{\infty}nP_{h}(n)-\frac{\theta}{y}\right)
-\mu_{2}^{c}\left( \sum_{n=1}^{\infty}nP_{c}(n)-\frac{1-\theta}{y}\right).
\nonumber
\end{eqnarray}
\noindent Since the $f$ dependence of $F_{chain}$ is due only to 
$F_{el}$, (\ref{Rdef}) yields $R$ as given by (\ref{Rdef2}). As before, 
$P_{h}(n)$ and $P_{c}(n)$ are determined by minimizing $F_{chain}(f)$ with
respect to these probabilities subject to two constraints: (i) The
normalization of the probabilities $P_{h}(n)$ and $P_{c}(n)$ and (ii) the
average sizes $k_{c}=(1-\theta )/y$ and $k_{h}=\theta /y$ of the coil and
helical domains. Upon substituting the resulting $P_{h}(n)$ and $P_{c}(n)$
in $F_{chain}$ we obtain $F_{chain}(f)$ as a function of $\theta $ and $y$. 
$\theta$ and $y$ are then determined by the equilibrium conditions $\partial
F_{chain}/\partial \theta =\partial F_{chain}/\partial y=0$. The details of
the calculation involved are described in Appendix E. $\theta $ and $y$ for
a given tension $f$ are

\begin{eqnarray}
\theta  &=&\frac{\sigma \,(1-\theta -y)\,sA\,\sinh (\gamma x)}{\gamma
x\left( 1-sAe^{\gamma x}\right) \left( 1-sAe^{-\gamma x}\right) }, \\
y &=&\frac{\sigma \,(1-\theta -y)}{2\gamma x}\,\ln \left( \frac{
1-sAe^{-\gamma x}}{1-sAe^{\gamma x}}\right) ,
\end{eqnarray}

\noindent where

\begin{equation}
A(x,\theta,y) = \frac{(1-\theta-y) \, x}{(1-\theta) \, \sinh (x)}.
\end{equation}

\noindent The probabilities $P_{c}(n)$ and $P_{h}(n)$ are

\begin{eqnarray}
P_{c}(n) &=&\frac{y}{1-\theta -y}\left( \frac{1-\theta -y}{1-\theta }
\right)^{n}, \\
P_{h}(n) &=&\frac{\sigma }{y}\,(1-\theta -y)\,\left( sA\right)^{n}\frac{
\sinh (n\gamma x)}{n\gamma x}.
\end{eqnarray}

\noindent $P_{h}(n)$, as $P_{c}(n)$, is essentially an exponentially
decaying function. However, the characteristic decay constants are now
dependent on $f$. The dependence is explicit for $P_{h}(n)$ while for 
$P_{c}(n)$ the $f$ dependence is implicit, {\it via} $\theta $ and $y$.

\begin{figure}[tbp]
\begin{center}
\epsfig{file=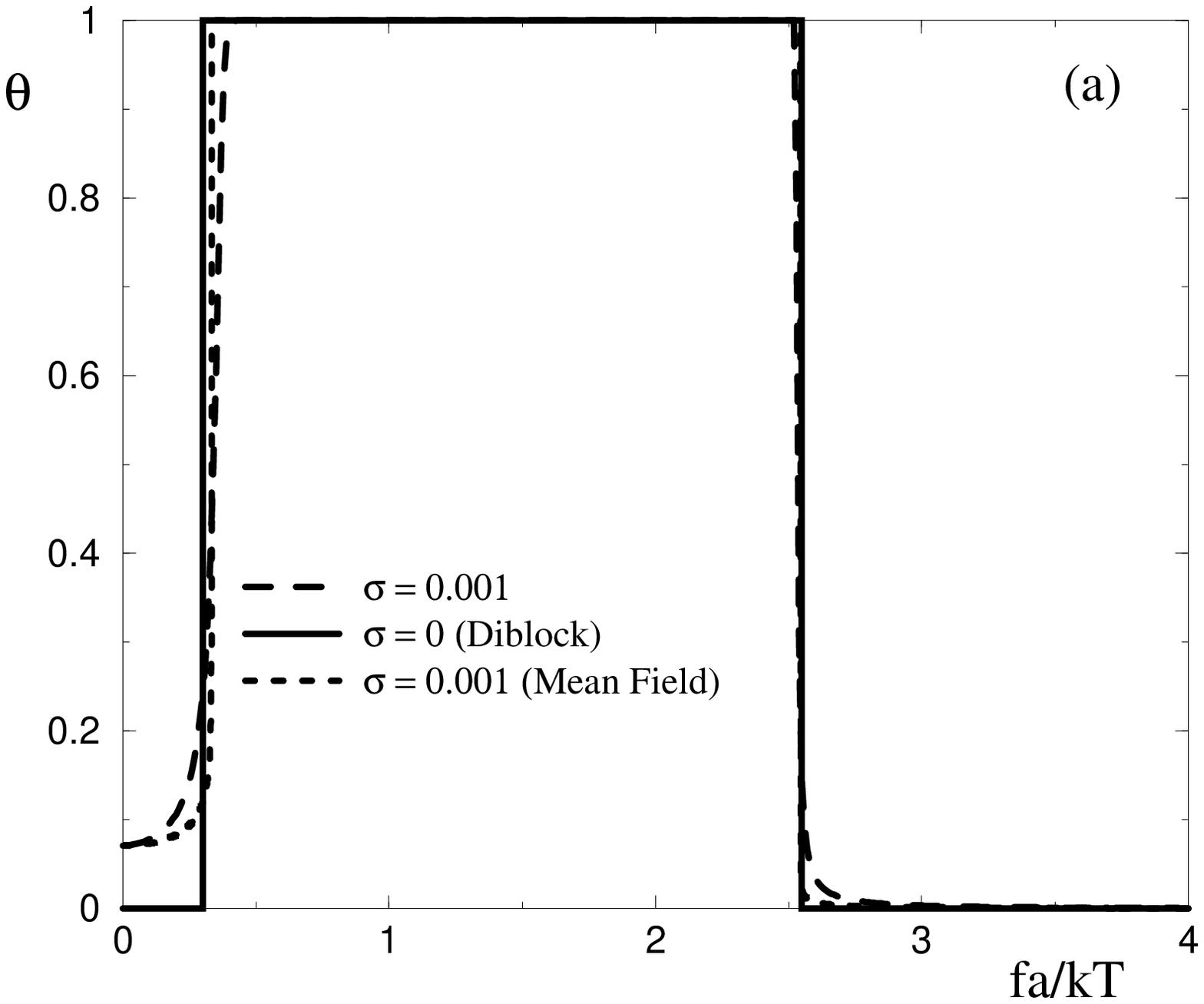,width=3in}
\epsfig{file=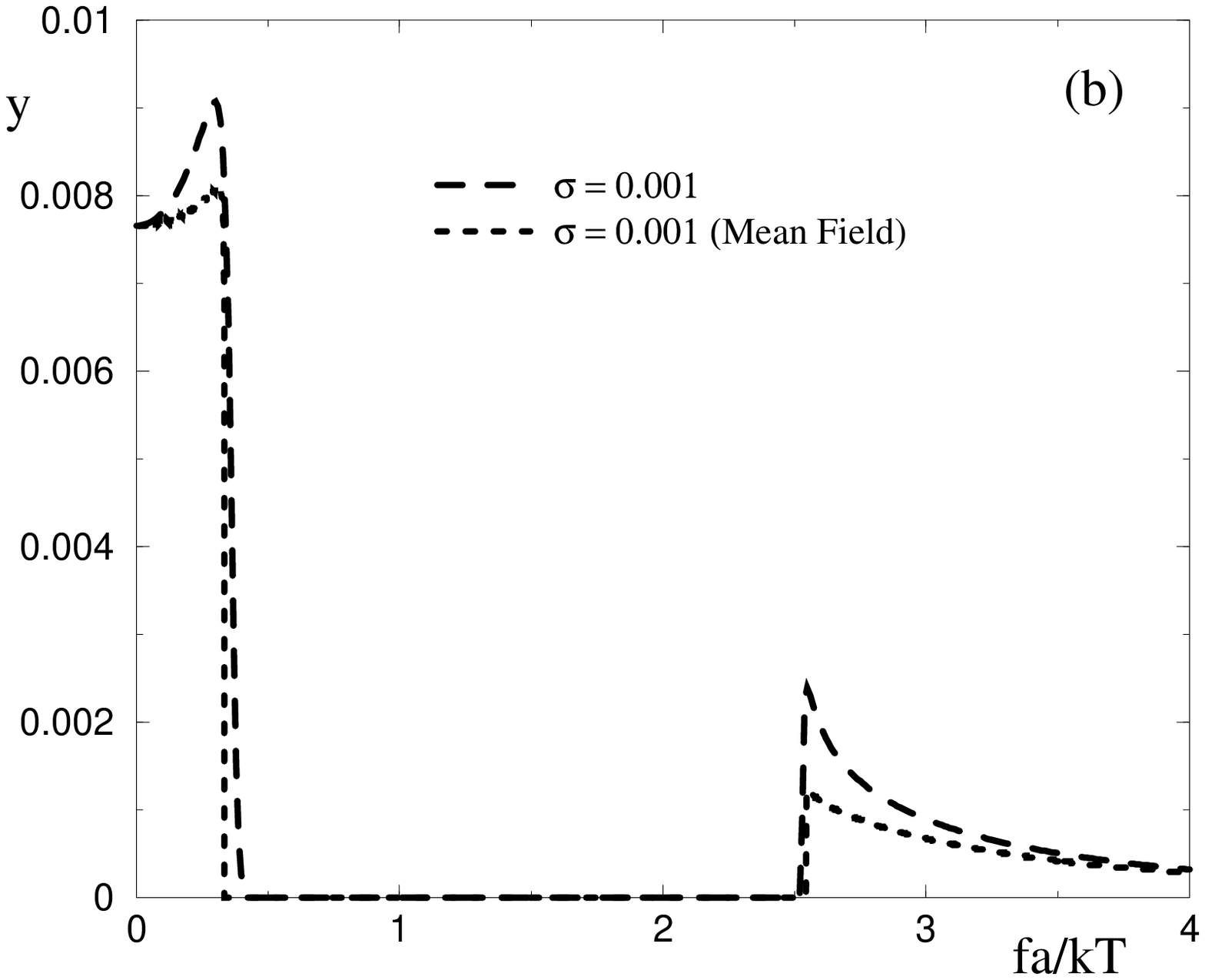,width=3in}
\caption{\label{theta2}
Plots of $\theta$ (a) and $y$ (b) {\em vs.} $fa/kT$ as
obtained from the rigorous solution (dashed lines), 
the $S_{mix} = 0$ or diblock approximation (continuous line) 
and the Mean Field approximation (dotted line) of the ``rigid 
helices'' model for $s=0.9 $ and $\gamma = 2/5$.}
\end{center}
\end{figure}

In Fig.~\ref{theta2}, the equilibrium $\theta $ and $y$, 
are plotted as functions of 
$x=fa/kT$ for different values of $\sigma $. Clearly, the results of the 
$S_{mix}=0$ or diblock approximation are approached in the limit of strong
cooperativity, $\sigma \rightarrow 0$. In particular, $y$ approaches zero
while $\theta $ exhibits a step like behavior. Thus, $\theta \simeq 0$
outside the plateaus range, $x<x_{1co}=0.30$ and $x>x_{2co}=2.55$, while
within the plateaus range, $x_{1co}<x<x_{2co}$ we find $\theta \simeq 1$ and 
$y\simeq 0$. Consequently, the mean size of the helical domains, 
$k_{h}=\theta /y$, becomes very large compared to $P$. In turn, this suggests
that the validity of the model of helical domains as rigid rods is
questionable. Clearly, the rigid helices model is meaningful only while the
persistence length is much larger than $k_{h}$.

Once the equilibrium characteristics of the chain for a given $f$ are found,
it is possible to obtain the force-extension diagram. The reduced end-to-end
distance, $r$, as a function of the reduced tension, $x$, is (Appendix E)

\begin{equation}
r=(1-\theta ){\cal L}(x)+\theta \gamma \left[ \coth (\gamma x)-\frac{sA}{
\sinh (\gamma x)}-\frac{y}{\theta \gamma x}\right] .
\end{equation}

\begin{figure}[tbp]
\begin{center}
\epsfig{file=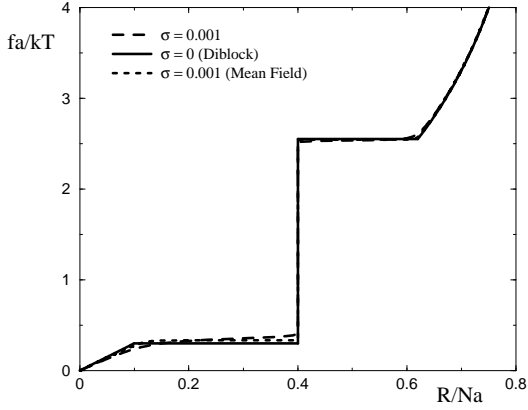,width=3in}
\caption{\label{force}
The extension force law as obtained from the rigorous solution
(dashed lines), the $S_{mix}=0$ or diblock approximation (continuous 
line) and the Mean Field approximation (dotted line) of the
``rigid helices'' model for $s=0.9$ and $\gamma =2/5$.}
\end{center}
\end{figure}

\noindent The corresponding force-extension diagram is plotted in 
Fig.~\ref{force}.
Again, the results of the $S_{mix}=0$ or diblock approximation are
approached in the $\sigma \rightarrow 0$ limit. Note the ``jump'' in the
force at $R=N\gamma a$~\cite{foot1}. At this point, between the two
plateaus, the polypeptide is in a ``purely'' helical state, $\theta \simeq 1$
and $y\simeq 0$. This jump is another indication of difficulties due to the
approximation of the helices as rigid rods. As we shall see in the next
section, this unphysical feature disappears when the finite persistence
length of the helices is allowed for.

It is of interest to compare the exact solution of the model to the mean
field approximation (MF)~\cite{Buhot,Mario}. The MF approximation
involves the removal of the explicit coupling between $P_{c}(n)$, $P_{h}(n)$
and $f$. As noted earlier, this decoupling is exact for\ $P_{c}(n)$. For 
$P_{h}(n)$ the decoupling is achieved by replacing $y\sum_{n}P_{h}(n)n\gamma 
{\cal L}(n\gamma x)$ in (\ref{Rdef2}) by $\theta \gamma {\cal 
L}(\theta \gamma x/y)$. In effect, all helical domains are assigned an equal
size of $k_{h}=\theta /y$ and the effect of their polydispersity on $F_{el}$
is ignored. The corresponding free energy is

\begin{eqnarray}
\frac{F_{chain}^{MF}}{NkT} &=&-\theta \ln s-y\ln \sigma +(\theta -y)\ln 
\frac{\theta -y}{\theta }  \label{eq_free_MF} \\
&&+y\ln \frac{y}{\theta } +(1-\theta -y)\ln \frac{1-\theta -y}{1-
\theta }+y\ln \frac{y}{1-\theta } \nonumber \\
&&-(1-\theta ){\cal L}_{int}(x)-y{\cal L}_{int}(\theta \gamma x/y). 
\nonumber
\end{eqnarray}
\noindent The equilibrium conditions $\partial F_{chain}/\partial
\theta =$ $\partial F_{chain}/\partial y=0$ lead to

\begin{eqnarray}
&&\frac{(1-\theta )(\theta -y)}{\theta (1-\theta -y)} =\frac{sx}{\sinh (x)}
\exp \left[\gamma x {\cal L}\left(\frac{\theta \gamma x}{y}\right) 
\right] ,\\
&& \frac{y^{2}}{(\theta -y)(1-\theta -y)} = \\
\nonumber & & \hskip 0.4cm  = \frac{\sigma y}{\theta \gamma x}
\sinh \left( \frac{\theta \gamma x}{y}\right) \exp \left[ -\frac{\theta
\gamma x}{y}{\cal L}\left( \frac{\theta \gamma x}{y}\right) \right] .
\end{eqnarray}

\noindent The overall features of the MF solution are rather similar
to the exact solution (Fig.~\ref{force}). 
The two primary differences are: (i)
The MF solution indicates that the transition between the plateaus involves
a second order phase transition~\cite{transition} in marked distinction to
the exact solution. (ii) Because of (i) the plateaus do not exhibit an
inflection point and the characteristic value of the slope, at $r=\gamma $,
scales as $\sigma^{2}$\cite{Mario}.

\subsection{Flexible helices}

The assumption of perfectly rigid helical sequences implies an infinite
persistence length. This assumption is reasonable so long as the size of the
helical domains is small compared to the known values of their persistence
length, $P$: A typical value is $P=2000\AA $ corresponding to $n_{h}
= P/a_h \sim 10^3$ monomers. As we have seen, the rigid rod approximation 
for an infinite
chain leads to helical domains that are larger. In turn, this leads to an
unphysical jump, located at $r=\gamma$, in the force-extension curve.

Two approaches allow to circumvent this difficulty. One approach treats each
helical monomer as an extendable spring {\em i.e.}, it introduces an elastic
penalty associated with the deviation of the length of the helical monomer
from its equilibrium length $a_{h}$. This approach was utilized in the
analysis of the extension behavior of DNA~\cite{Chatenay} and of
poly-(ethylene glycol)~\cite{Kreuzer}. In the following we adopt a
different approach that proved successful in fitting the force curves of
DNA. Within this approach the helical domains are considered as worm like
chains (WLC) instead of rigid rods (Appendix D).

In the WLC model the helical domains are viewed as semiflexible chains
undergoing undulations. The angular correlations along the chain decay
because of the thermal undulations. Within this model the decay length of
the angular correlations, $\lambda ,$ depends on the tension, $\lambda
=\lambda (f).$ For weakly stretched chains $\lambda $ is identical to the
persistence length, $P$, of the unperturbed chain, $\lambda
(f)=P$ while for stronger tensions $\lambda (f)=P(kT/fP)^{1/2}$ and the
correlations decay faster. A long helical domain of length $n\gamma a$ may
thus be considered as a FJC of $n\gamma a/2\lambda (f)$ ``rigid'' segments
of length $2\lambda (f)$ and the reduced end-to-end distance is 
\begin{equation}
r=(1-\theta ){\cal L}(x)+\theta \gamma {\cal L}(\alpha )  \label{dist_flex}
\end{equation}
\noindent where ${\cal L}(x)$ is the Langevin function and $\alpha =\alpha
(n_{h}\gamma x)=2f\lambda (f)/kT$~\cite{Kuhn}. 
As before, the expression for $r$
reflects two contributions. The first term corresponds to the alignment of
the coil monomers with the applied tension using the FJC model. The second
term describes the alignment of the semiflexible helical domains. In the
following we assume that the helical domains are larger than $\lambda (f)$
that is, $\lambda (f)\ll n\gamma a$. The error introduced by this assumption
diminishes as $f$ increases and $\lambda (f)$ decreases. Within this
approximation each $2\lambda (f)$ segment in a helical domain is aligned
independently. This is in marked contrast to the rigid helices model where
all the monomers within a helical domain align as a unit. As a result we may
utilize $k_{h}=\theta /y=\sum_{n}nP_{h}(n)$ in order to obtain 
$y\sum_{n=1}^{\infty }P_{h}(n)\,n\gamma \,{\cal L}(\alpha )=\theta \gamma 
{\cal L}(\alpha )$. Consequently, $r$ is independent of the probabilities 
$P_{c}(n)$ and $P_{h}(n)$. Expression (\ref{dist_flex}) for $R(f)$ enables us
to obtain (Appendix D) the corresponding elastic free energy 
$F_{el}(f)=-\int_{0}^{f}Rdf^{\prime }$ 
\begin{equation}
\label{eq_Fel}
F_{el}/NkT=-(1-\theta ){\cal L}_{int}(x)-\frac{\theta }{4n_{h}}\left(
4n_{h}\gamma x-\alpha \right) {\cal L}\left( \alpha \right) .
\end{equation}
\noindent In marked distinction to the ``rigid helices'' scenario, this
elastic free energy is independent of the probabilities $P_{h}(n)$ and 
$P_{c}(n)$. Accordingly, $P_{h}(n)$ and $P_{c}(n)$ are not modified
explicitly by the extension and obey (\ref{proba_h}) and (\ref{proba_c}). In
other words, the mean field approximation provides an exact solution of the
problem. Altogether, the free energy of a polypeptide, using the WLC model
for the helical domains, is 
\begin{eqnarray}
&&\frac{F_{chain}(f)}{NkT} =-\theta \ln s-y\ln \sigma +(\theta -y)
\ln \frac{ \theta -y}{\theta }  \label{eq_free_flex} \\
&&+y\ln \frac{y}{\theta } +(1-\theta -y)\ln \frac{1-\theta -y}{1-
\theta }+y\ln \frac{y}{1-\theta} \nonumber \\
&&-(1-\theta ){\cal L}_{int}(x)-\frac{\theta }{4n_{h}}[4n_{h}\gamma 
x-\alpha ]{\cal L}(\alpha ).  \nonumber
\end{eqnarray}
\noindent This expression is similar to $F_{chain}$ for a free, undeformed
chain (\ref{eq_free1}). The two free energies differ in two respects: (i) 
(\ref{eq_free_flex}) contains an additive term $-{\cal L}_{int}(x)$ which is
independent of $\theta $ and $y$. (ii) The force independent $s$ in (\ref
{eq_free1}) is replaced by a force dependent $\tilde{s}(x)$ 
\begin{equation}
\tilde{s}(x)=\frac{sx}{\sinh (x)}\exp \left[ \left( \gamma x-\alpha
/4n_{h}\right) {\cal L}(\alpha )\right] .  \label{stilde}
\end{equation}
\noindent Consequently, the equilibrium conditions $\partial
F_{chain}/\partial \theta =\partial F_{chain}/\partial y=0$ lead to
expressions of the same form as those found for the free chains but with 
$\tilde{s}=\tilde{s}(x)$ replacing $s$ 
\begin{eqnarray}
\theta  &=&\frac{1}{2}+\frac{1}{2}\,\frac{\tilde{s}-1}{\sqrt{(\tilde{s}
-1)^{2}+4\,\sigma \tilde{s}}}, \\
y &=&\frac{2\,\sigma \tilde{s}\,\left[ (\tilde{s}-1)^{2}+4\,\sigma \tilde{s}
\right] ^{-1/2}}{\tilde{s}+1+\sqrt{(\tilde{s}-1)^{2}+4\,\sigma \tilde{s}}}.
\end{eqnarray}
\noindent Substitution of the equilibrium values of $\theta $ and $y$ for a
given $x$ in (\ref{dist_flex}) yields the corresponding equilibrium
end-to-end distance.

\begin{figure}[tbp]
\begin{center}
\epsfig{file=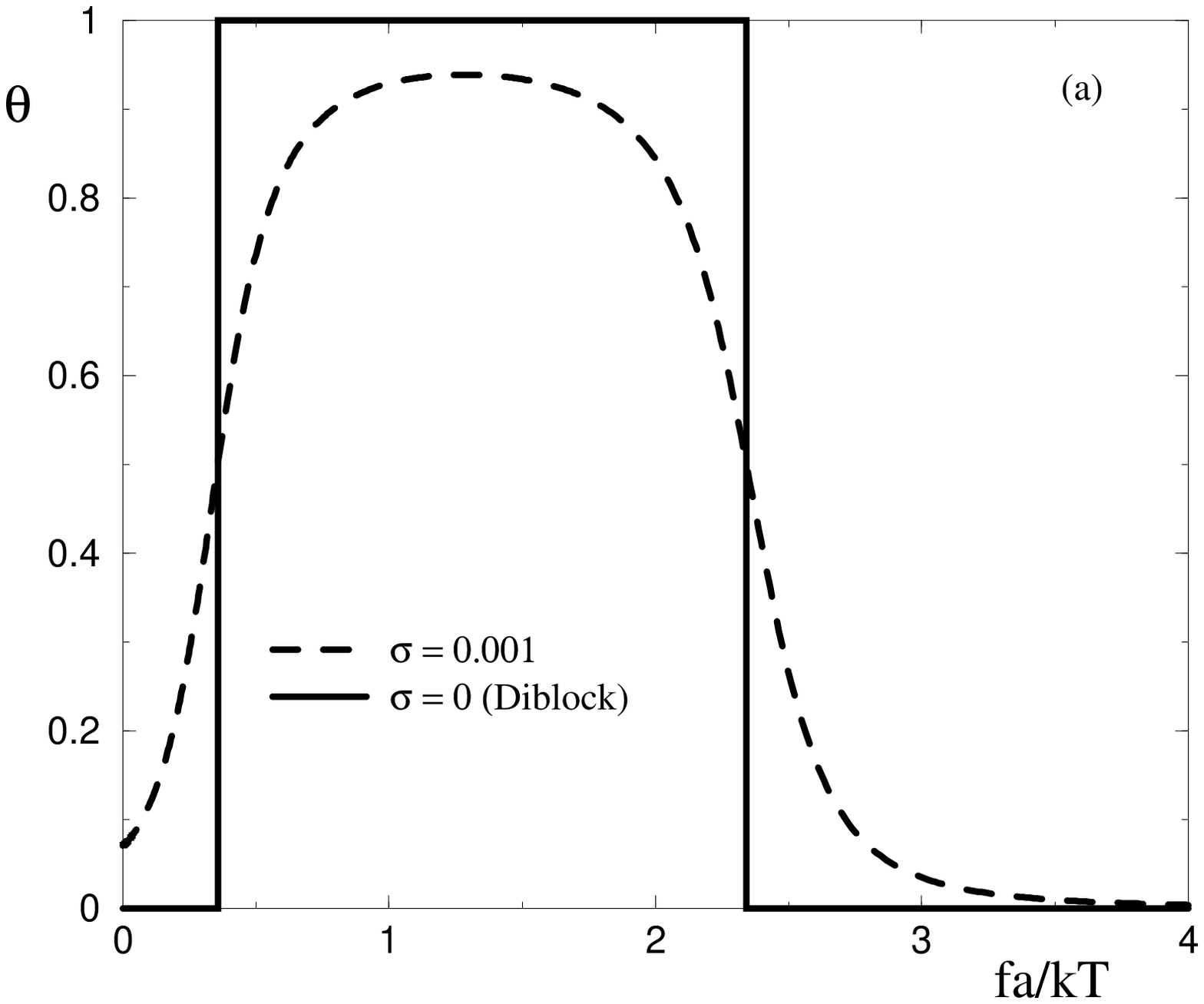,width=3in}
\epsfig{file=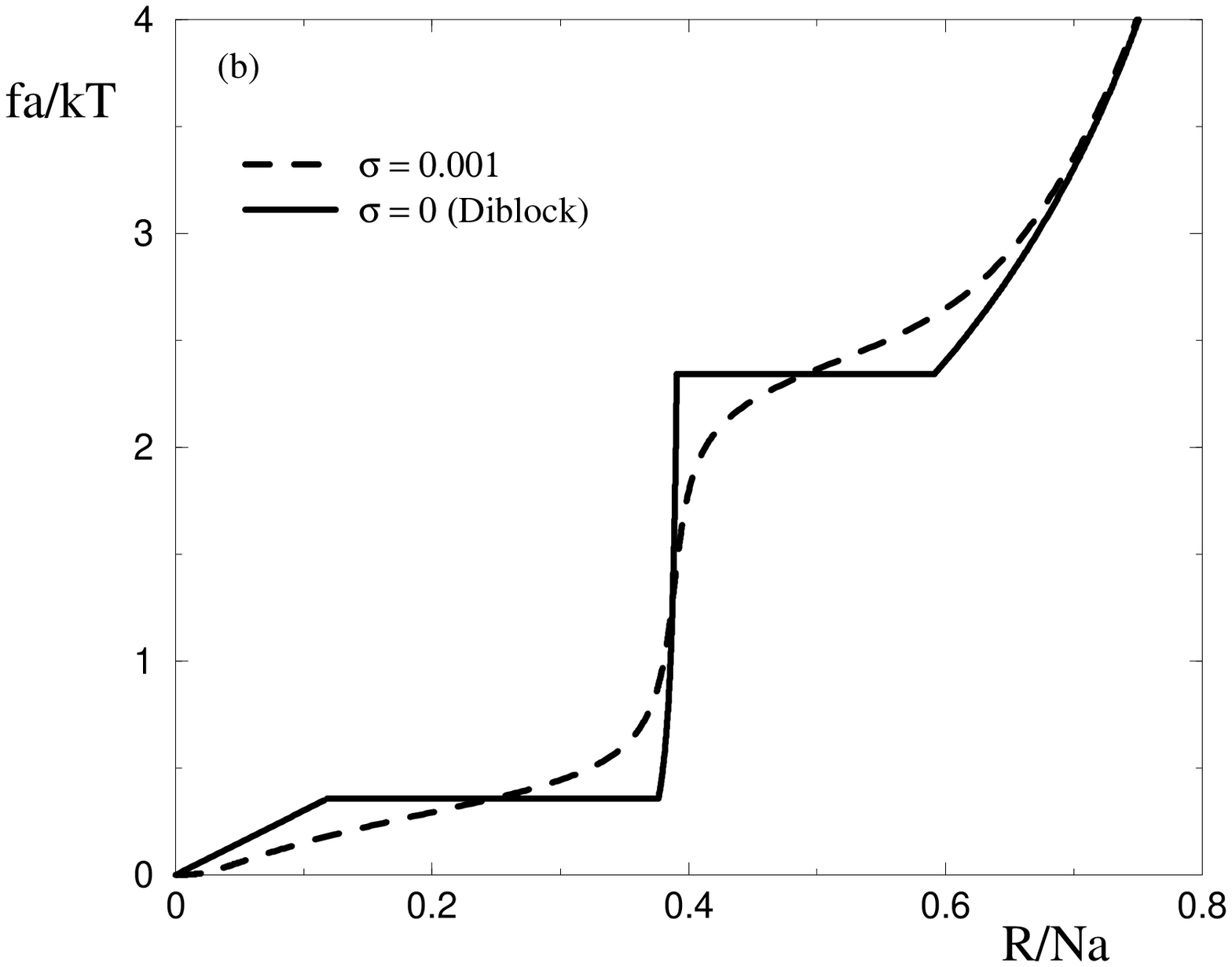,width=3in}
\caption{\label{thetaWLC}
Plots of the helical fraction $\theta$ {\em vs.} the
reduced force $fa/kT$ (a) and plots of the reduced force $fa/kT$ {\em vs.} 
$r=R/Na$ (b) as obtained from the rigorous solution (dashed lines) and the 
$S_{mix}=0$ approximation (continuous line) of the ``semiflexible helices''
model for $s=0.9$, $\gamma = 2/5$ and $n_h = 1000$.}
\end{center}
\end{figure}

Plots of $\theta $ {\em vs.} $fa/kT$ and $fa/kT$ {\em vs.} $R/Na$, as
obtained for the ``semiflexible helices'' scenario for different $\sigma $,
are depicted in Fig.~\ref{thetaWLC}. 
The replacement of the ``rigid helices'' by
semiflexible ones leads to the disappearance of the ``jump'' in the force
law. Another difference with respect to Fig.~\ref{theta2}a 
is that $\theta =1$ is never
attained. Again, we recover the appropriate diblock approximation in the
limit of $\sigma \rightarrow 0$.

All the force-extension curves in Fig.~\ref{thetaWLC}b 
cross at three different points.
The external points correspond to a reduced force $x_{\pm }$ satisfying 
$\tilde{s}(x_{\pm })=1$. In turn, this condition is equivalent to the
equilibrium condition $\partial F_{chain}/\partial \theta =0$ for $F_{chain}$
corresponding to the appropriate diblock approximation in which the coil is
modeled as FJC and the helix as a WLC~\cite{footnote3}. 
In particular, $x_{\pm }$ correspond
to points in the plateaus where $\theta =1/2$, $r_{\pm }=\left[ {\cal L}%
(x_{\pm })+\gamma {\cal L}(\alpha _{\pm })\right] /2$ and $\alpha _{\pm
}\equiv \alpha (n_{h}\gamma x_{\pm })$. The inner crossing point corresponds
to $x_{0}$ for which $\tilde{s}(x)$ is maximal. $x_{0}$ satisfies the
condition ${\cal L}(x_{0})=\gamma {\cal L}\left[ \alpha (n_{h}\gamma x_{0})
\right] $ {\em i.e.}, to an equality between the projected length of the
coil and the helical monomers along the force. From (\ref{dist_flex}) and 
(\ref{stilde}), it is apparent that this condition is independent of 
$\sigma $. The plateaus depicted in Fig.~\ref{thetaWLC}b exhibit a 
small but finite slope. Its characteristic value, at the inflection 
points $x_{\pm }$, scales with $\sigma^{1/2}$~\cite{footnoteX}.

\section{Discussion and conclusion}

Our analysis of the extension behavior of polypeptides utilized two models.
The two differ in the description of the helical domains. In one model, the
helical domains are viewed as perfectly rigid and unextendable rods. The
``rigid helices'' model is reasonable so long as the length of the helical
domains is small compared to the persistence length. In the second approach
the helical domains are treated as long semiflexible polymers within the
worm like chain (WLC) model. This approach is preferable when the length of
the helical domains is larger than the force dependent decay length $\lambda
(f)$. For each of the models, we presented a rigorous solution and two
approximations. Within the diblock approximation one neglects the mixing
entropy associated with the polydispersity of the domains. In the mean field
approximation there is no direct coupling between the distribution of sizes
of the domains and the applied tension. In all cases considered, our
analysis is based on the minimization of a phenomenological free energy.
This approach is equivalent to the transfer matrix method but it is
physically transparent and mathematically simple.

The general features of the extension behavior of the polypeptides are
independent of the model and the approximation scheme. Three different
scenarios are possible for the extension of long polypeptides: (a) For 
$T>T_{c}$ the polypeptide remains in a coil configuration for all extensions.
(b) In the range $T^{\ast }<T<T_{c}$ the chain is initially in a coil state
but it undergoes a helix-coil transition upon extension. The helical domains
eventually melt upon stronger extension. Consequently, the force law
exhibits two plateaus associated with a helix-coil coexistence. (c) For 
$T<T^{\ast }$ the unstretched polypeptide is a helix which melts upon
stretching thus leading to a force law with a single plateau.

In the second scenario, $T^{\ast }<T<T_{c}$, it is possible to distinguish
between five regimes: (I) For weak extensions the polypeptide behaves as a
random coil exhibiting a linear response. (II) Stronger extensions, and the
associated loss of configurational entropy, favor the formation of helical
domains. The associated coexistence of helical and coil domains gives rise
to a lower plateau. (III) A steep crossover corresponding to the stretching
of a stiff, mainly helical, polymer leads to (IV) a second, higher plateau
due to the break-up of the helices when the imposed end-to-end distance is
larger then the length of a fully helical chain. This plateau is also
associated with helix-coil coexistence. (V) When the helical content is
reduced to zero the chain exhibits a strong extension behavior of a random
coil, with deviations from the linear response behavior due to finite
extensibility effects.

\begin{figure}[tbp]
\begin{center}
\epsfig{file=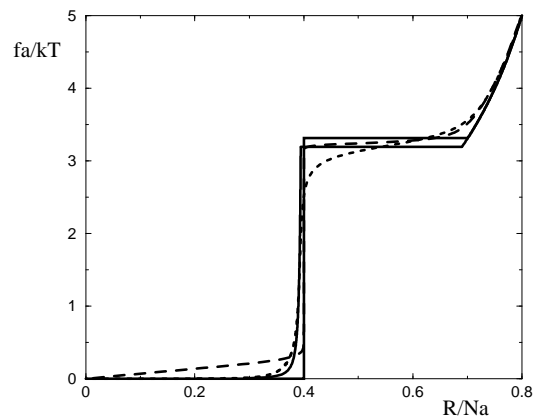,width=3in}
\caption{\label{f_s1.1}
Plots of the reduced force $fa/kT$ {\em vs.} the reduced end-to-end
distance $R/Na$ for the $s>1$ scenario ($s = 1.1$, $\gamma = 2/5$
and $n_h = 1000$). Long dashed line: ``rigid helices'' model 
($\sigma = 0.01$), dashed line: ``semiflexible helices'' model ($\sigma 
= 0.001$) and continuous line: corresponding diblock approximations.}
\end{center}
\end{figure}

In the third scenario, $T<T^{\ast }$, the regimes (I) and (II) disappear and
regime (III) extends down to $R=0$. The small slope of the force law in this
regime (Fig.~\ref{f_s1.1}) is reminiscent of a plateau. 
However, this quasi-plateau is
not associated with a helix-coil coexistence. Rather, it is due to the easy
alignment of the persistent segments in the helix.

The two models, ``rigid'' and ``semiflexible'' helices, differ with regard
to the precise details of the force diagram. Thus, region (III) occurs as a
``jump'' within the ``rigid helices'' model but is a steep yet gradual
increase in the ``semi-flexible helices'' model. The slope of
the plateaus within the rigid helices model scales as $\sigma \left[ \ln
\sigma +\ln |\ln \sigma |\right] ^{2}$ \cite{Mario} while in the 
``semi-flexible helices'' model it scales as $\sigma^{1/2}$. Overall,
the crossovers between the various regimes are sharper in the ``rigid
helices'' model.

The diblock approximation, with the appropriate elastic term, allows to
recover the main features, length and force scales, of each model. However,
within this $S_{mix}=0$ approximation the transition takes place as a first
order phase transition. As a result it yields perfect plateaus, $f=f_{1co}$
or $f=f_{2co}$, and sharp crossovers. The mean field approximation, where
the polydispersity is decoupled from the tension is exact for the 
``semiflexible helices'' model. In the case of the ``rigid helices''
model this approximation, while it captures the overall features of the
force law also gives rise to a number of artifacts~\cite{Mario}.

Neither of the models considered captures the physics completely. The
``rigid helices'' model fails to describe the behavior of helical domain
that are long compared to the persistence length. Within the ``semiflexible
helices'' model all helical domains are assumed to be long in comparison to 
$\lambda (f)$. Accordingly, this model is likely to mishandle
short helical domains that do behave as rigid rods. The importance of this
deficiency diminishes however as $f$ increases and $\lambda (f)$ decreases.
A full description of the system requires a model combining the positive
features of the two models described above.

While our analysis focused on the case of homopolypeptides forming $\alpha$
-helices, our approach can be extended to more complicated systems. In the
case of heteropolypeptides it is necessary to allow for the residue
dependence of $\Delta f$. For double stranded DNA it is necessary to allow
for long range interactions as given by the Stockmayer formula for 
$\sigma$~\cite{Rouzina1,Rouzina2,Rouzina3,Rouzina4}.

\subsection*{Acknowledgment}

We would like to thank M. N. Tamashiro and P. Pincus as well as I. Rouzina 
and V. A. Bloomfield for providing us with their results prior to 
publication. The work of AB was supported by the Marie Curie Fellowship
HPMF-CT-1999-00328.

\appendix

\section{Probability of sizes}

The minimization of the free energy (\ref{eq_free2}) with respect to the
probabilities $P_{h}(n)$ and $P_{c}(n)$ yields

\begin{eqnarray}
y \ln P_{h}(n) &=& \mu_{1}^{h} - y + n \mu_{2}^{h}, \\
y \ln P_{c}(n) &=& \mu_{1}^{c} - y + n \mu_{2}^{c}.
\end{eqnarray}

\noindent These are solved by $P_{h}(n) = A_{h} B_{h}^{n}$ and $P_{c}(n) =
A_{c} B_{c}^{n}$ where

\begin{eqnarray}
A_{h(c)} &=& \exp \left( \frac{\mu_{1}^{h(c)}-y}{y}\right), \\
B_{h(c)} &=& \exp \left( \frac{\mu_{2}^{h(c)}}{y}\right).
\end{eqnarray}

\noindent The parameters $A_{h}$, $A_{c}$, $B_{h}$ and $B_{c}$ are
determined using the constraints

\begin{eqnarray}
1 &=& \sum_{n=1}^{\infty} P_{h}(n) = \frac{A_{h} B_{h}}{1-B_{h}}, \\
\frac{\theta}{y} &=& \sum_{n=1}^{\infty} n P_{h}(n) = \frac{A_{h} 
B_{h}}{(1-B_{h})^{2}}, \\
1 &=& \sum_{n=1}^{\infty} P_{c}(n) = \frac{A_{c} B_{c}}{1-B_{c}}, \\
\frac{1-\theta}{y} &=& \sum_{n=1}^{\infty} n P_{c}(n) = \frac{A_{c} 
B_{c}}{(1-B_{c})^{2}},
\end{eqnarray}

\noindent yielding $A_{h} = y/(\theta-y)$, $B_{h} = (\theta-y)/\theta$, 
$A_{c} = y/(1-\theta-y)$ and $B_{c} = (1-\theta-y)/(1-\theta)$ thus leading
to the expressions (\ref{proba_h}) for $P_{h}(n)$ and (\ref{proba_c}) for 
$P_{c}(n)$.

\section{Fractions of helices and of helical monomers}

The equilibrium conditions $\partial F_{0}/\partial \theta = \partial
F_{0}/\partial y = 0$ with $F_{0}$ given by (\ref{eq_free1}) lead to

\begin{eqnarray}
(1-\theta )(\theta-y) &=& s \, \theta \, (1-\theta -y),  
\label{appendixa1} \\
y^{2} &=& \sigma \, (\theta-y)(1-\theta-y).  \label{appendixa2}
\end{eqnarray}

\noindent From (\ref{appendixa2}), we obtain

\begin{equation}
y=\frac{1}{2\alpha }\left[ \sqrt{1+4\,\alpha \,\theta \,(1-\theta )}-1\right]
\end{equation}

\noindent or

\begin{equation}
y = \frac{2 \, \theta (1-\theta)}{\sqrt{1 + 4 \, \alpha \, \theta (1-\theta)}
+ 1}
\end{equation}

\noindent where $\alpha = (1-\sigma)/\sigma$. The second condition 
(\ref{appendixa1}) yields

\begin{equation}
y = \frac{\theta \, (1-\theta)(s-1)}{\theta \, (s+1)-1}.
\end{equation}

\noindent Combining these two equations allows us to eliminate $y$ and
obtain an equation for $\theta$ as a function of $s$ and $\sigma$,

\begin{equation}
\sqrt{1 + 4 \, \alpha \, \theta \, (1-\theta)} + 1 = \frac{2 \, \theta \,
(s+1) - 2}{s-1}.
\end{equation}

\noindent leading to

\begin{eqnarray}
\theta & = & \frac{1}{2} + \frac{1}{2} \frac{s-1}{\sqrt{ (s-1)^2 + 4 \,
\sigma s}}, \\
y & = & \frac{2 \, \sigma s \, ((s-1)^2 + 4 \, \sigma s)^{-1/2}}{s + 1 + 
\sqrt{(s-1)^2 + 4 \sigma s}}.
\end{eqnarray}

\section{The Freely Jointed Chain Model}

The freely jointed chain model may be considered as the macromolecular
analog of the Langevin theory of paramagnetism~\cite{Hill}. A flexible chain
comprising $N$ monomers is the counterpart of $N$ non interacting magnetic
dipoles. The monomer length $a$ is the analog of the magnetic moment $\mu $,
the applied tension ${\bf f}$ plays the role of the magnetic field ${\bf B}$
and the induced magnetic moment, $M$ is the analog of the end to end
distance $R.$ Each monomer is assigned an orientational energy 
$-fa\cos \phi$ where $\phi $ is the angle between the force and the 
segment. The reduced
end-to-end distance $R/Na$ of the polymer, at a temperature $T$, is the mean
value of $\cos \phi $:

\begin{equation}
\frac{R}{Na}=\langle \cos \phi \rangle _{T}=\frac{\displaystyle{
\int_{0}^{2\pi }\cos \phi e^{x\cos \phi }\sin \phi d\phi }}{
\displaystyle{
\int_{0}^{2\pi }e^{x\cos \phi }\sin \phi d\phi }}={\cal L}(x)
\end{equation}
\noindent where $x=fa/kT$ is the reduced force and ${\cal L}(x)=\coth x-1/x$
is the Langevin function. The corresponding elastic free energy in the fixed 
$f$ ensemble, as given by $F_{el}(f)=-\int_{0}^{f}Rdf'$, leads to
the following free energy per monomer: 
\begin{equation}
\frac{F_{el}(f)}{NkT}=-{\cal L}_{int}(x)
\end{equation}
\noindent where ${\cal L}_{int}(x)=\ln \left[ \sinh x/x\right] $, is the
integrated Langevin function.

The elastic free energy in the fixed $R$ ensemble $F_{el}(R)=
\int_{0}^{R} f dR' = Rf+F_{el}(f)$
is the Legendre transform of $F_{el}(f)$ where $r=R/Na$ and $x$ are related
by $r={\cal L}(x)$. For weak extensions $F_{el}(R)$ reduces to the familiar
Gaussian form $F_{el}(R) \simeq 3kTR^{2}/{2Na^{2}}$ while for strong 
extensions it diverges as $F_{el} (R) \sim \ln f\sim -\ln (1-r)$.

\section{The Wormlike Chain Model}

Within the wormlike chain model (WLC) the polymer is viewed as a bendable 
rod~\cite{DE}. In the following we consider the case of a chain of constant
contour length $L$ and a single bending modulus $E=kTP$ where $P$ is the
persistence length of the unperturbed chain. The chain trajectory is
described by the position of a point on the chain, ${\bf R}(s),$ at the
contour length $s.$\ The unit tangent vector ${\bf u}(s)=\partial {\bf R}/
\partial s$ specifies the local orientation of the chain. A straight rod
corresponds to ${\bf u}(s)=const^{\prime }$ or $\partial {\bf u}/\partial
s=0.$ Accordingly, the bending energy is a quadratic function of 
$\partial {\bf u} /\partial s$. Since ${\bf u} \cdot \partial 
{\bf u}/\partial s=0$, the only quadratic form is 
\begin{equation}
U_{bend}=\frac{E}{2}\int_{0}^{L}ds\left( \frac{\partial 
{\bf u}}{\partial s} \right)^{2}
\end{equation}
and the conformational distribution of the chain, its partition function, 
up to a contour length $s$, is 
\begin{equation}
\Psi \propto \exp \left[ -\frac{P}{2}\int_{0}^{s}ds' \left( 
\frac{\partial {\bf u}}{\partial s'}\right)^{2}\right] 
\end{equation}
where $\Psi$ is a function of ${\bf u}(0) = {\bf u}_{0}$ and ${\bf u}
(s)$ {\em i.e,} $\Psi = \Psi ({\bf u}_{0}, {\bf u}(s))$. This
describes a Gaussian process with the constraint ${\bf u}^{2}=1$. 
As such it is analogous to rotational diffusion and obeys 
$\frac{\partial \Psi }{\partial s}=\frac{1}{2P}\widehat{\Re }^{2} \Psi$ 
where $\widehat{\Re }\equiv {\bf u}\times \frac{\partial }{\partial 
{\bf u}}$ is the rotational operator
that is related to the angular momentum operator in quantum mechanics, 
$\widehat{L}$, as $-i\widehat{\Re }=\widehat{L}$. The correlations along the
unperturbed chain decay as $\langle {\bf u}(s)\cdot {\bf u}(0)\rangle =\exp
(-s/P).$ When the chain is subjected to an external field $U_{ext}({\bf u)}$
the expression for $\Psi $ becomes 
\begin{equation}
\Psi \propto \exp \left[ -\frac{P}{2} \int_{0}^{s} ds' \left( \frac{
\partial {\bf u}}{\partial s'} \right)^{2}-\frac{1}{kT}
\int_{0}^{s} ds' U_{ext}({\bf u)}\right] 
\end{equation}
The effect of tension ${\bf f}$ is specified by $U_{ext}({\bf u})= - {\bf
f} \cdot {\bf u}=-f \cos \phi $ thus leading to 
\begin{equation}
\label{eq_psi}
\frac{\partial \Psi }{\partial s}=\left( \frac{1}{2P}\widehat{\Re }^{2}+
\frac{f\cos \phi }{kT}\right) \Psi 
\end{equation}
It is possible to find eigenstate satisfying $\frac{\partial \psi }{\partial
s}=-g\psi$. For a long chain the free energy is extensive and the
eigenfunction expansion of $\Psi ({\bf u}_0, {\bf u}(L))$ involves terms of 
the form $\exp (-gL)\psi ({\bf u}_{0})\psi ({\bf u}(L))$ where $gkT$ is 
the free energy per unit length due to the bending fluctuations.

An analytical solution of equation (\ref{eq_psi}) is not available. 
Marko and Siggia~\cite{Marko} argued that the WLC free energy is 
determined by the lowest $g$ in the spectrum and utilized a variational 
approach in order to determine $g$. The force law is then recovered from 
$R/L = - kT d g/d f$. To this end they chose the trial function 
$\psi ({\bf u})\propto \exp \left[
\alpha \cos (\phi )/2\right]$ where $\alpha $ is a variational
parameter and $\phi$ is the angle between ${\bf u}$ and ${\bf f}$. 
As we shall see later, $\alpha = 2f\lambda /kT$ where $\lambda
=\lambda (f)$ is a force dependent decay length that replaces $P$ in
characterizing the decay of $\langle {\bf u}(s)\cdot {\bf u}(0) \rangle$.
Assuming $\int d{\bf u}^{2}$ $\psi ^{2}({\bf u})=1$ this leads 
to~\cite{footF}
\begin{equation}
\label{eq_g} ga = \min_{\alpha} Ga =  
\min_{\alpha }\left( \frac{\alpha a}{4P}-x\right) {\cal L}(\alpha )
\end{equation}
\noindent where ${\cal L}(x)=\coth x-1/x$ is the Langevin function and 
$x=fa/kT$ is the reduced force. The argument $\alpha = \alpha
(n_{h}\gamma x)$ that minimizes $g$ is specified by 
$\partial G/\partial \alpha = 0$ which leads to:
\begin{equation}
\label{eq_alpha}
{\cal L}(\alpha )=[4n_{h}\gamma x-\alpha ]{\cal L}^{\prime }(\alpha ).
\end{equation}
\noindent  where ${\cal L}^{\prime }(x)=d{\cal L}(x)/dx$. 
Here $n_{h}\gamma x=Pf/kT$. The end-to-end distance 
is simply deduced from :
\begin{equation}
\label{RWLC}
R = -L \frac{dga}{dx} = - L \frac{\partial Ga}{\partial x}
- L \frac{\partial Ga}{\partial \alpha} \frac{\partial \alpha}{\partial x}
= L {\cal L}(\alpha ).  
\end{equation}
\noindent This end-to-end distance is identical to the one obtained from
the FJC model if the reduced force $2n_{h}\gamma x$~\cite{Kuhn} is replaced
by $\alpha (n_{h}\gamma x)$. For small forces, when ${\cal L}(\alpha
)\approx \alpha /3$ and ${\cal L}^{\prime }(\alpha )\approx 1/3$ 
(\ref{eq_alpha}) leads to $\alpha (n_{h}\gamma x)\approx 2n_{h}\gamma x$ or 
$\lambda \approx P$. For high tension $\alpha $ increases more slowly. 
In this limit ${\cal L} (\alpha )\approx 1-1/\alpha$ while 
${\cal L}^{\prime }(\alpha )\approx 
1/\alpha ^{2}$ and (\ref{eq_alpha}) yields $\alpha (n_{h}\gamma x)\simeq 
(4 n_{h}\gamma x)^{1/2}$ or $\lambda \approx P(kT/fP)^{1/2}$.
The difference between the
variational approximation and the exact solution of the WLC model is less
than $1.5\%$ over the whole range of forces. The associated elastic free
energy, $F_{el}(f)= - \int_{0}^{f}Rdf'$ is
\begin{equation}
F_{el}(f)=LkTg
\end{equation}
since $Rdf=-LkTdg$ and where $g$ corresponds to the explicit 
expression in (\ref{eq_g}) with $\alpha = \alpha(n_h \gamma x)$ 
according to (\ref{eq_alpha}). In Eq. (\ref{eq_Fel}) this elastic 
free energy is used for a helical domain of length $L = n 
\gamma a$ with $n$ monomers.

\section{Rigid helices approximation}

In this appendix we determine the probabilities $P_c (n)$ and $P_h (n)$ of a
coil and a helical segment with $n$ monomers as well as the fractions of
helical monomers $\theta$ and of helices $y$.

The extremum condition of $F_{chain}$ (\ref{eq_free_rigid}) with respect to 
$P_{c}(n)$ and $P_{h}(n)$ yields

\begin{eqnarray}
y \ln P_c (n) & = & \mu_1^c - y + n \mu_2^c, \\
y \ln P_h (n) & = & \mu_1^h - y + n \mu_2^h + y \, {\cal L}_{int} (n \gamma
x).
\end{eqnarray}

\noindent This leads to

\begin{eqnarray}  
\label{eq_distri_c}
P_c (n) & = & \exp \left( \frac{\mu_1^c-y}{y} + n \, \frac{\mu_2^c}{y}
\right), \\
\label{eq_distri_h}
P_h (n) & = & \exp \left( \frac{\mu_1^h - y}{y} + n \, \frac{\mu_2^h}{y}
\right) \frac{\sinh (n \gamma x)}{n \gamma x}.
\end{eqnarray}

The probability $P_{c}(n)$ is easily determined since it is not modified by
the stretching. The normalization of the probability $P_{c}(n)$ and the mean
size $k_{c} = (1-\theta)/y = \sum_{n} n P_{c}(n)$ lead to (Appendix A)

\begin{eqnarray}
\exp \left( \frac{\mu_1^c-y}{y} \right) & = & \frac{y}{1-\theta-y}, \\
\exp \left( \frac{\mu_2^c}{y} \right) & = & \frac{1-\theta-y}{1-\theta}
\end{eqnarray}

\noindent and:

\begin{equation}  
\label{prob_Pc}
P_c (n) = \frac{y}{1-\theta-y} \left(\frac{1-\theta-y}{1-\theta} \right)^n.
\end{equation}

The probability $P_{h}(n)$ is modified by the stretching and is more
difficult to obtain. To do so we consider the two remaining equilibrium
conditions $\partial F_{chain}/\partial \theta = 0$ and $\partial
F_{chain}/\partial y = 0$ after substitution of the equilibrium form of 
$P_{c}(n)$ as given by (\ref{prob_Pc})

\begin{eqnarray}  
\label{lns}
\ln s & = & - \ln \left( \frac{1-\theta-y}{1-\theta} \right) + 
\frac{\mu_2^h}{y} + {\cal L}_{int} (x), \\
\label{lnsigma}
\ln \sigma & = & \ln \left( \frac{y}{1-\theta-y} \right) +
\sum_{n=1}^{\infty} P_h (n) \ln P_h (n) \\
& & - \frac{\mu_2^h \theta}{y^2} - \sum_{n=1}^{\infty} P_h (n) {\cal L}
_{int} (n \gamma x).  \nonumber
\end{eqnarray}

\noindent Using (\ref{eq_distri_h}) for $P_{h}(n)$ we may rewrite 
(\ref{lnsigma}) as

\begin{equation}  
\label{lnsigma2}
\ln \sigma =\ln \left(\frac{y}{1-\theta-y} \right) + \frac{\mu_1^h}{y} - 1.
\end{equation}

\noindent From (\ref{lnsigma2}) and (\ref{lns}) we obtain

\begin{eqnarray}
\exp \left( \frac{\mu _{1}^{h}-y}{y}\right) &=&\frac{\sigma 
(1-\theta -y)}{y}, \\
\exp \left( \frac{\mu _{2}^{h}}{y}\right) &=&\left( \frac{1-\theta 
-y}{1-\theta }\right) \frac{sx}{\sinh x}.
\end{eqnarray}

\noindent Thus leading to

\begin{equation}
P_{h}(n)=\frac{\sigma \,(1-\theta -y)^{n+1}}{y\,(1-\theta )^{n}}\left( 
\frac{ s\,x}{\sinh x}\right)^{n}\frac{\sinh (n\gamma x)}{n\gamma x}.
\end{equation}

$\theta$ and $y$ are obtained using the normalization condition for 
$P_{h}(n) $ and the constraint on the average size of the helical domains 
$k_{h} = \theta/y$

\begin{eqnarray}
\sum_{n=1}^{\infty }\frac{(1-\theta -y)^{n+1}}{(1-\theta )^{n}}\left( 
\frac{s\,x}{\sinh x}\right) ^{n}\frac{\sinh (n\gamma x)}{n\gamma x} 
&=&\frac{y}{\sigma }, \\
\sum_{n=1}^{\infty }\frac{(1-\theta -y)^{n+1}}{(1-\theta )^{n}}\left( 
\frac{s\,x}{\sinh x}\right) ^{n}\frac{\sinh (n\gamma x)}{\gamma x} 
&=&\frac{\theta }{\sigma }.
\end{eqnarray}

\noindent Using the definition $\sinh (x)=(e^{x}-e^{-x})/2$ as well as the
relations

\begin{eqnarray}
\sum_{n=1}^{\infty} X^n & = & \frac{X}{1-X}, \\
\sum_{n=1}^{\infty} \frac{X^n}{n} & = & - \ln (1-X),
\end{eqnarray}

\noindent we obtain

\begin{eqnarray}
\frac{y}{\sigma } &=&\frac{1-\theta -y}{2\gamma x}\,\ln \left( 
\frac{1-sAe^{-\gamma x}}{1-sAe^{\gamma x}}\right) ,  \label{y} \\
\frac{\theta }{\sigma } &=&\frac{(1-\theta -y)\,sA\sinh (\gamma x)}{\gamma
x\,\left( 1-sAe^{\gamma x}\right) \left( 1-sAe^{-\gamma x}\right) },
\label{theta}
\end{eqnarray}

\noindent where

\begin{equation}
A(x,\theta,y) = \frac{(1-\theta-y) \, x}{(1-\theta) \, \sinh (x)}.
\end{equation}

\noindent These two equations for $\theta $ and $y$ can be easily solved
numerically. To this end we introduce a new variable $\epsilon =(1-\theta
-y)/(1-\theta )$. Equation (\ref{y}) may then be rewritten as

\begin{equation}
1-\epsilon = \frac{\sigma \epsilon}{2 \gamma x} \ln \left[ \frac{1 - s
A(x,\epsilon) e^{-\gamma x}}{1 - s A(x,\epsilon) e^{\gamma x}}\right]
\end{equation}

\noindent with $A(x,\epsilon) = \epsilon x/\sinh (x)$. This equation allows
to determine $\epsilon$ as a function of $s$, $\sigma$, $\gamma$ and $x$. 
$\theta$ is then obtained from (\ref{theta}) in the form

\begin{equation}
\frac{\theta}{1-\theta} = \frac{\sigma \epsilon s A(x,\epsilon) \sinh
(\gamma x)}{\gamma x [1 - s A(x,\epsilon) e^{-\gamma x}] [1 - s
A(x,\epsilon) e^{\gamma x}]}.
\end{equation}

\noindent Knowing $\epsilon$ and $\theta$, $y$ is simply determined from $y
= (1-\epsilon)(1-\theta)$.

To obtain the end-to-end distance $R$ as a function of the tension $f$ we
use $R=-\partial F_{chain}/\partial f$ which leads to 
\begin{equation}
r=(1-\theta ){\cal L}(x)+y\sum_{n=1}^{\infty }P_{h}(n)\,n\gamma 
\,{\cal L} (n\gamma x).
\end{equation}
\noindent The second term may be tackled exactly using the definition 
${\cal L}(x)=\coth x -1/x$: 
\begin{eqnarray}
r &=&(1-\theta ){\cal L}(x)+y\sum_{n=1}^{\infty }P_{h}(n)\,n\gamma 
\,\frac{\cosh (n\gamma x)}{\sinh (n\gamma x)}-\frac{y}{x}  \nonumber \\
&=&(1-\theta ){\cal L}(x)+\frac{\sigma }{2x}(1-\theta -y)\left( \Lambda
_{+}+\Lambda _{-}\right) -\frac{y}{x}
\end{eqnarray}

\noindent with

\begin{equation}
\Lambda_{\pm} = \frac{s A(x,\theta,y) e^{\pm \gamma x}}{1 - s A(x,\theta,y)
e^{\pm \gamma x}}.
\end{equation}

The expression for $r$ may be simplified using (\ref{theta}):

\begin{eqnarray}
\Lambda _{+} + \Lambda _{-} &=& \frac{2 s A(x,\theta,y) \cosh (\gamma x) - 2
s^{2} A^{2}(x,\theta,y)}{[1 - s A(x,\theta,y) \, e^{\gamma x}][1 - s
A(x,\theta,y) \, e^{-\gamma x}]},  \nonumber \\
&=& \frac{2 \theta \gamma x [\cosh (\gamma x) - s A(x,\theta,y)]}{ \sigma
(1-\theta-y) \sinh (\gamma x)}
\end{eqnarray}

\noindent which yields to

\begin{equation}
r = (1-\theta) {\cal L}(x) + \theta \gamma \left[ \coth (\gamma x) - \frac{s
A(x,\theta,y)}{\sinh (\gamma x)}\right] - \frac{y}{x}.
\end{equation}

\end{multicols}

\begin{references}
\bibitem{footnote}  To be precise, this approach is equivalent to a transfer
matrix description utilizing a second order transfer matrix. In turn, this
description is sufficient for the analysis of the helix-coil transition in
an infinite chain.
\bibitem{estimate}  It is customary to estimate $\Delta h$ by the energy of 
$H$-bond, $E_{H}$. The values of $E_{H}$ vary in the range of $10-20kJ/mol$.
For $E_{H} = -10kJ/mol$ we find $\sigma \approx 3.3 \times 10^{-4}$
at room temperature~\cite{Grosberg}.
\bibitem{constraint}  The constraint that helical domains incorporate at
least three monomers can be easily introduced by changing the lower boundary
of sums involving $P_{h}(n)$ from $n=1$ to $n=3$.
\bibitem{smix}  The final expression for the mixing entropy for the free
chain can be calculated directly~\cite{Buhot}. $S_{mix}=k\ln W$ where $W$ 
is the product of two factors: $W_{th}={N\theta \choose Ny}$ 
the number of possible placements of $Ny$ boundaries to separate 
$N\theta$ helical bonds into $Ny$ helical domains and $W_{tc}=
{N(1-\theta ) \choose Ny}$
the number of ways to place $Ny$ boundaries to separate $N(1-\theta )$
coil bonds into $Ny$ coil domains. Using the Sterling formula $N!\sim N\ln
N-N$ we obtain $S_{mix}(\theta ,y,N)=k\ln W\simeq Nk\left[ \theta s\left( 
\frac{y}{\theta }\right) +(1-\theta )s\left( \frac{y}{1-\theta }\right) 
\right]$ where $s(x)$ stands for $-x\ln x-(1-x)\ln (1-x)$.
\bibitem{footnote2}  In the fixed $f$ ensemble, the free energy is 
$F_{chain}/NkT =-\theta \ln s-(1-\theta ){\cal L}_{int}(x)-
\gamma \theta x$ where ${\cal L}_{int}(x)=\ln \left[ \sinh (x)/x\right]$ 
(see Appendix C) and the boundaries of the plateau are specified by the
equilibrium condition $\partial F_{chain}/\partial \theta =0=-\ln s+
{\cal L}_{int}(x_{\pm })-\gamma x$ while the end-to-end distance is: 
$R=-\partial F_{chain}/\partial f=Na[(1-\theta ) {\cal L}(x)+\theta 
\gamma ]$.
\bibitem{footnote3}  In this case the $S_{mix}=0$ approximation leads to
(see Appendix D and section IVB) $F_{chain}/NkT=-\theta \ln
s-(1-\theta ){\cal L}_{int}(x)-\frac{\theta }{4n_{h}}[4n_{h}\gamma x-\alpha]
{\cal L}(\alpha)$. The domain boundaries are specified by $0=-\ln
s+{\cal L}_{int}(x_{\pm })-[4n_{h}\gamma x_{\pm }-\alpha _{\pm }]{\cal L}
(\alpha _{\pm })/4n_{h}$ with $\alpha _{\pm }\equiv \alpha (n_{h}\gamma
x_{\pm })$ and the end-to-end distance is: $R=Na[(1-\theta ){\cal L}(x)+
\theta \gamma {\cal L}(\alpha )]$.
\bibitem{foot1}  A careful study of the behavior at this apparent jump shows
that there is no first order transition.
\bibitem{transition}  Even though the force law displays a jump, the phase
transition is a second order transition since the order parameter $\theta $
changes continuously and attains unity.
\bibitem{footnoteX}  To characterize the slope we expand $\tilde{s}(x)$
around $x_{\pm }$ where $\tilde{s}(x_{\pm })=1$. In the immediate vicinity
of $x_{\pm }$ we have $\tilde{s}(x)\simeq 1+\delta (x-x_{\pm })$ where 
$\delta$ is a constant. Accordingly, $\theta \simeq \frac{1}{2}\left( 1+
\frac{\tilde{s}-1}{\sqrt{4\,\sigma \tilde{s}}}\right) \simeq \frac{1}{2}
\left[ 1+\frac{\delta (x-x_{\pm })}{2\,\sigma^{1/2}}\right]$
and $r=r_{\pm }+\frac{\delta (x-x_{\pm})}{2\,\sigma ^{1/2}} 
\left\{ {\cal L}(x_{\pm })-\gamma {\cal L}[\alpha (x_{\pm })]\right\}$
leading to $x-x_{\pm }\sim \sigma ^{1/2}(r-r_{\pm })$.
\bibitem{footF}  The variational eigenstate $\psi = M \exp \left[ 
\alpha \cos (\phi)/2 \right]$ where $M$ is the normalization factor. 
With $d^2 u = \frac{1}{4\pi}
 d \omega  \sin \phi \, d\phi$, the normalization condition $\int d^{2}u 
\, \psi^{2}=1$ yields $M^{2} = \alpha /\sinh (\alpha )$. 
Since there is no $\omega$ dependence $\widehat{\Re }=i
\overrightarrow{\varphi } \partial /\partial \theta $ where 
$\overrightarrow{\varphi }$ is the appropriate unit vector. 
Consequently (\ref{eq_psi}) leads to $g =  
\int d^2 u \left\{\frac{1}{2P} ( \widehat{\Re }\psi)^{2} - \frac{f}{kT} 
\psi^2 \cos \phi \right\} = \frac{M^{2}}{2} \int_{-1}^{1} \left\{ 
\frac{\alpha^2}{8 P} (1-y^{2}) \exp (\alpha y) - \frac{f}{kT}  y \exp 
(\alpha y) \right\} dy.$
\end{references}
\end{document}